\documentclass[amsmath,amssymb,prb,superscriptaddress,reprint,showpacs,longbibliography]{revtex4-1}
\usepackage{graphicx}
\usepackage{dcolumn}
\usepackage[colorlinks=true,linkcolor=blue,citecolor=blue,urlcolor=blue]{hyperref}
\usepackage[usenames,dvipsnames]{xcolor}
\usepackage{soul}
\usepackage{braket}
\usepackage{bm}
\usepackage{upgreek}
\usepackage{nicefrac}

\newcommand{\MC}[1]{\mathcal{#1}}

\newcommand{\ud}{\mathrm{d}}
\newcommand{\iu}{\mathrm{i}}
\newcommand{\Tr}{\mathrm{Tr}}
\newcommand{\IM}{\mathrm{Im}}

\newcommand{\EF}{{E_{\text{F}}}}

\newcommand{\VEC}[1]{\mathbf{#1}}

\newcommand{\PHIR}{\phi_{\text{R}}}
\newcommand{\NORM}[1]{\lvert #1 \rvert}

\begin{document}

\title{Anatomy of magnetic anisotropy induced by Rashba spin-orbit interactions}

%%%%%
\author{Gaurav Chaudhary}
\email{gaurav@physics.utexas.edu}
\affiliation{Department of Physics, The University of Texas at Austin, Austin, Texas 78712, USA}
%%%%%
\author{Manuel dos Santos Dias}
\affiliation{Peter Gr\"{u}nberg Institut and Institute for Advanced Simulation, Forschungszentrum J\"{u}lich \& JARA, 52425 J\"{u}lich, Germany}
%%%%%
\author{Allan H. MacDonald}
\affiliation{Department of Physics, The University of Texas at Austin, Austin, Texas 78712, USA}
%%%%%
\author{Samir Lounis}
\email{s.lounis@fz-juelich.de}
\affiliation{Peter Gr\"{u}nberg Institut and Institute for Advanced Simulation, Forschungszentrum J\"{u}lich \& JARA, 52425 J\"{u}lich, Germany}

\date{\today}


%%%%%%%%%%%%%%%%%%%%%%%%%%%%%%%%%%%%%%%%%%%%%%%%%%%%%%%%%%%%%%%%%%%%%%%%%%%%%%%%%%%%%%%%%%%%%%%%%%%%
\begin{abstract}
Magnetic anisotropy controls the orientational stability and switching properties of magnetic states,
and therefore plays a central role in spintronics.  
First-principles density-functional-theory calculations are able, in most cases, to provide a satisfactory description of bulk and interface contributions to the magnetic anisotropy of particular film/substrate combinations.  
In this paper we focus on achieving a simplified understanding of some trends in interfacial magnetic anisotropy 
based on a simple tight-binding model for quasiparticle states in a heavy-metal/ferromagnetic-metal bilayer film.  
We explain how to calculate the magnetic anisotropy energy of this model
from the quasiparticle spin-susceptibility, compare with more conventional approaches using either a 
perturbative treatment of spin-orbit interactions or a 
direct calculation of the dependence of the energy on the orientation of the magnetization,
and show that the magnetic anisotropy can be interpreted as a competition between a Fermi-sea term favoring 
perpendicular anisotropy and a Fermi-surface term favoring in-plane anisotropy.
Based on this finding, we conclude that perpendicular magnetic anisotropy 
should be expected in an itinerant electron thin film when the spin magnetization density is
larger than the product of the band exchange splitting and the Fermi level density-of-states of the magnetic state. 
\end{abstract}
%%%%%%%%%%%%%%%%%%%%%%%%%%%%%%%%%%%%%%%%%%%%%%%%%%%%%%%%%%%%%%%%%%%%%%%%%%%%%%%%%%%%%%%%%%%%%%%%%%%%

\maketitle

%%%%%%%%%%%%%%%%%%%%%%%%%%%%%%%%%%%%%%%%%%%%%%%%%%%%%%%%%%%%%%%%%%%%%%%%%%%%%%%%%%%%%%%%%%%%%%%%%%%%
\section{Introduction}

Spintronics\cite{Zutic2004} aims to utilize the electron spin as the active degree of freedom for information 
storage and processing.  Bilayers containing an interface~\cite{Hellman2017} between a thin film of a heavy-metal
and a magnetic one are important hybrid materials in spintronics, as they combine magnetic
order, strong spin-orbit interactions, and broken inversion symmetry.
Strong spin-orbit coupling (SOC) derived from the heavy-metal layer and 
inversion symmetry broken by the interface, 
combined with exchange interactions  of the magnetic layer, can lead to 
perpendicular magnetic anisotropy~\cite{Johnson1996} (PMA), 
Dzyaloshinskii-Moriya interactions\cite{Dzyaloshinskii1957,Moriya1960}, 
spin-orbit torques, Rashba-Edelstein effects, and more.~\cite{Manchon2015}
Spin-orbit interactions near the interface provide a handle to alter these properties
by tuning chemical composition, interface structure, 
or gate voltages, as demonstrated most extensively
for magnetic anisotropy.\cite{Chiba2008,Maruyama2009}

Magnetic anisotropy energy (MAE) 
refers to the dependence of the total energy of a magnetic system 
on the real-space orientation of its magnetization.
The MAE is responsible for the orientational stability of magnetic domains, and
hence lies at the heart of both magnetic hard disk drives and 
magnetic random access memories.
There are two main contributions to the MAE\cite{Skomski2008}: the 
magnetocrystalline anisotropy which arises from electronic spin-orbit 
interactions, and shape anisotropy which arises from the magnetostatic dipolar interaction.
For a thin ferromagnetic film, the magnetostatic energy is minimized when the 
magnetization is in the plane of the film, leading to in-plane magnetic anisotropy (IMA).
To stabilize PMA, the magnetocrystalline anisotropy energy must overcome the shape anisotropy.
From the technological point of view, PMA is very important, since it
enables an increased bit storage density, through a reduced size of the 
magnetic domains that store each bit of information.
For this reason, considerable experimental and theoretical effort has been
devoted to the design, growth and understanding of magnetic materials 
displaying PMA.

The fact that spin-orbit coupling contributes to the MAE
was pointed out by Bloch\cite{Bloch1931} and van Vleck\cite{vanVleck1937},
and Brooks\cite{Brooks1940} first outlined its description in terms of the underlying
electronic structure.  More recently, Bruno\cite{Bruno1989} pointed to 
an appealing perturbative connection between MAE and the anisotropy of the orbital magnetic moment, 
which was later generalized in Ref.~\onlinecite{Laan1998}.
Density functional theory (DFT) calculations for transition-metal 
systems\cite{Daalderop1991,Daalderop1994} showed that Bruno's connection holds for 3d transition metals and their compounds, and also for thin films,
although some counter-examples have been uncovered recently, both experimentally and theoretically\cite{Andersson2007}.
Total energy differences from self-consistent DFT calculations provide a 
reliable but cumbersome way of computing the MAE for a specific 
target system~\cite{Halilov1998}, and can be simplifed by use of 
the magnetic force theorem~\cite{Weinert1985,Daalderop1991}.
A different approach is to evaluate directly the derivative of the energy with 
respect to the ferromagnetic orientation, the so-called torque 
method\cite{Wang1996}.  Recently, it has been proposed by 
Antropov \emph{et al.}~\cite{Antropov2014a} that a numerically stable way of computing the MAE is to evaluate half of the anisotropy in the SOC energy term in the Hamiltonian, adapting to electronic structure calculations an 
idea already advanced by van der Laan.\cite{Laan1999}

The one-band Rashba model\cite{Bychkov1984a} is often used
to illustrate the effects of SOC on band structure and materials properties related
to surfaces and interfaces. It can describe the interplay between SOC
and the coupling of the electron spin to a magnetic condensate, 
in particular to interpret the properties of 
magnetic/heavy-metal bilayers\cite{Manchon2015}. Ref.~\onlinecite{Barnes2014} presented a simple theoretical description of the MAE using the free-electron Rashba model, and pointed out that the finite bandwidth must be taken into account, as confirmed in Ref.~\onlinecite{Kim2016a}.
A gate voltage was experimentally demonstrated to control the Rashba coupling 
strength\cite{Caviglia2010}, which might provide a route to the electrical 
control of the MAE\cite{Chiba2008}.

In this work we study the magnetocrystalline anisotropy of a ferromagnet/heavy-metal bilayer, driven by interfacial Rashba SOC, highlighting different physical regimes and considering different ways of interpreting the results.
We develop the theory for the finite bandwidth case, employing the tight-binding approximation.
The behavior of the MAE is analyzed with respect to the three competing energy scales: the non-relativistic kinetic energy $t'$, the Rashba SOC strength $t''$, and the strength of the exchange coupling $J$ to the ferromagnetic order parameter.
We contrast the global definition of the MAE (energy difference between different ferromagnetic directions of the system) with its local definition (curvature of the energy for a given ferromagnetic direction).
This curvature of the energy is evaluated from the electronic spin susceptibility, providing a new way to compute the MAE.
We show that the Fermi surface contribution favors IMA, while the Fermi sea contribution favors PMA.
This indicates that both the overall band filling and the relative contributions from individual bands
at a fixed total filling play an important role in stabilizing PMA.
The analytic treatment of the half-filled case provides a figure of merit for PMA in this model, and numerical calculations recover the IMA $\rightarrow$ PMA $\rightarrow$ IMA behavior of the MAE when the filling is increased from zero to two electrons\cite{Kim2016a}.
The recent proposal that the 
MAE is half of the anisotropy in the SOC energy is also explored.
We consider three qualitatively different parameter regimes for
 detailed study: (i) strong exchange ($J \gg t' \gg t''$), (ii) intermediate exchange ($J \sim t' \gg t''$), and (iii) weak exchange ($t' \gg J \sim t''$).

The paper is organized as follows.
In Sec.~\ref{sec:model} we present the tight-binding model and the theoretical and numerical methods, and illustrate the main features of the electronic structure.
Different ways of computing the MAE and an overview of the results are discussed in Sec.~\ref{sec:mae}, connecting to previous work.
The half-filled case is analyzed using perturbation theory in Sec.~\ref{sec:halffilling}, where we 
prove that it always has PMA.  This analytic calculation suggests 
a useful figure of merit for MAE.
Then the MAE is studied in detail in Sec.~\ref{sec:casestudies}, focusing on 
the three physically distinct cases mentioned above.
Our conclusions are gathered in Sec.~\ref{sec:conclusions}, and some 
derivations and analytical calculations are presented in three appendices.
%%%%%%%%%%%%%%%%%%%%%%%%%%%%%%%%%%%%%%%%%%%%%%%%%%%%%%%%%%%%%%%%%%%%%%%%%%%%%%%%%%%%%%%%%%%%%%%%%%%%

%%%%%%%%%%%%%%%%%%%%%%%%%%%%%%%%%%%%%%%%%%%%%%%%%%%%%%%%%%%%%%%%%%%%%%%%%%%%%%%%%%%%%%%%%%%%%%%%%%%%
\section{Model and methods}\label{sec:model}
%%%%%%%%%%%%%%%%%%%%%%%%%%%%%%%%%%%%%%%%%%%%%%%%%%%%%%%%%%%%%%%%%%%%%%%%%%%%%%%%%%%%%%%%%%%%%%%%%%%%
To illustrate the properties of itinerant electrons with broken inversion symmetry and 
SOC, we consider a two-dimensional square lattice with one 
orbital per site, nearest-neighbor hopping, 
and Rashba-like spin-momentum locking:
\begin{equation}
  \MC{H}_{\text{e}} = -\frac{1}{2} \sum_{\braket{i,j}} \sum_{s,s'} c_{is}^\dagger\Big(t'\sigma^0_{ss'} 
   - \iu\,t''\big(\hat{\VEC{z}} \times \hat{\VEC{R}}_{ij}\big) \cdot \bm{\upsigma}_{ss'}\Big) c_{js'} \quad .
\end{equation}
Here the sum is over near-neighbor links,
$c_{is}^\dagger$ and $c_{is}$ are the creation and annihilation operators for an electron with 
spin $s$ at a lattice site $\VEC{R}_i$, $\sigma^0$ is the unit $2\times2$ spin matrix, and $\bm{\upsigma} = (\sigma^x,\sigma^y,\sigma^z)$ is the vector of Pauli matrices.
The vector connecting site $i$ to site $j$ is $\VEC{R}_{ij} = \VEC{R}_j - \VEC{R}_i\,$, and the cross product
favors spin-orientations perpendicular to the bond direction,
$\hat{\VEC{R}}_{ij} = \VEC{R}_{ij} / \NORM{\VEC{R}_{ij}}$ and the 
normal to the lattice plane, $\hat{\VEC{z}}$.
The hopping strength is given by $t$, and the angle $\PHIR$ characterizes the relative strength of conventional spin-independent 
hopping $t' = 2 t \cos\PHIR$ and chiral Rashba hopping $t'' = 2 t \sin\PHIR$.

We impose Born-von Karman periodic boundary conditions and introduce the lattice Fourier transforms of the operators,
\begin{equation}
  c_{is} = \frac{1}{\sqrt{N}} \sum_{\VEC{k}} e^{\iu \VEC{k} \cdot \VEC{R}_i}\,c_{s}(\VEC{k}) \quad,
\end{equation}
\begin{equation}
  \frac{1}{N}\sum_{\VEC{k}} e^{\iu \VEC{k} \cdot (\VEC{R}_i-\VEC{R}_j)} = \delta_{ij}, \quad 
  \frac{1}{N}\sum_{i} e^{\iu (\VEC{k}' - \VEC{k}) \cdot \VEC{R}_i} = \delta_{\VEC{k}' \VEC{k}} \quad ,
\end{equation}
where $N$ is the number of lattice sites.

This leads to the $\VEC{k}$-space representation of the Hamiltonian matrix elements,
\begin{subequations}\label{eq:hrashba}
\begin{align}
  \MC{H}_{\text{e}}(\VEC{k}) &= \MC{H}_0(\VEC{k}) + \MC{H}_{\text{R}}(\VEC{k}) \quad, \\
  \MC{H}_0(\VEC{k}) &= -t'\left(\cos k_x + \cos k_y\right) \sigma^0 \quad, \\
  \MC{H}_{\text{R}}(\VEC{k}) &= - t''\left(\sin k_x\,\sigma^y - \sin k_y\,\sigma^x\right) \quad ,
\end{align}
\end{subequations}
where we have used the lattice constant as the unit of length.
For small $\VEC{k}$-vectors (setting $\hbar = 1$ and ignoring the leading constant term),
\begin{align}
  \MC{H}_{\text{e}}(\VEC{k}) &\approx \frac{t'(k_x^2 + k_y^2)}{2}\,\sigma^0 - t''\,\big(k_x\,\sigma^y - k_y\,\sigma^x\big) \nonumber\\
  &= \frac{k_x^2+k_y^2}{2m^*}\,\sigma^0 + \alpha\,(\VEC{k} \times \hat{\VEC{z}})\cdot\bm{\upsigma} \quad,
\end{align}
which is the form of the Hamiltonian for a Rashba electron gas, 
with $m^*$ the effective mass and $\alpha$ the Rashba parameter.

To model a ferromagnetic system, we add ferromagnetic exchange
between the quasiparticles and the magnetic condensate:
\begin{subequations}\label{eq:sdhamil}
\begin{align}
  \MC{H}(\VEC{k}) &= \MC{H}_{\text{e}}(\VEC{k}) - \VEC{B} \cdot \bm{\upsigma} \\
  &= \MC{H}_0(\VEC{k}) + \MC{H}_{\text{R}}(\VEC{k}) + \MC{H}_{\text{B}} \\
  &= E_0(\VEC{k})\,\sigma^0 - \VEC{b}(\VEC{k}) \cdot \bm{\upsigma} \quad ,
\end{align}
\end{subequations}
where
\begin{equation}
  E_0(\VEC{k}) = -t'\left(\cos k_x + \cos k_y\right) \quad,
\end{equation}
and
\begin{subequations}\label{eq:spinaxis}
\begin{align}
  \VEC{b}(\VEC{k}) &= \VEC{b}_{\text{R}}(\VEC{k}) + \VEC{B} \quad, \\
  \VEC{b}_{\text{R}}(\VEC{k}) &= t''\left(\sin k_y\,\hat{\VEC{x}} - \sin k_x\,\hat{\VEC{y}}\right) \quad , \\
  \VEC{B} &= J \left(\sin\theta\left(\cos\varphi\,\hat{\VEC{x}} + \sin\varphi\,\hat{\VEC{y}}\right) + \cos\theta\,\hat{\VEC{z}}\right) \quad .
\end{align}
\end{subequations}
Here $\VEC{b}_{\text{R}}(\VEC{k})$ is the Rashba spin-orbit field, and the coupling to the ferromagnetic background is given by $\VEC{B}$, where the spherical angles $\theta$ and $\varphi$ specify the magnetization orientation and $J$ is the strength of the coupling.

\begin{figure*}[!htb]
  \includegraphics[width=0.8\textwidth]{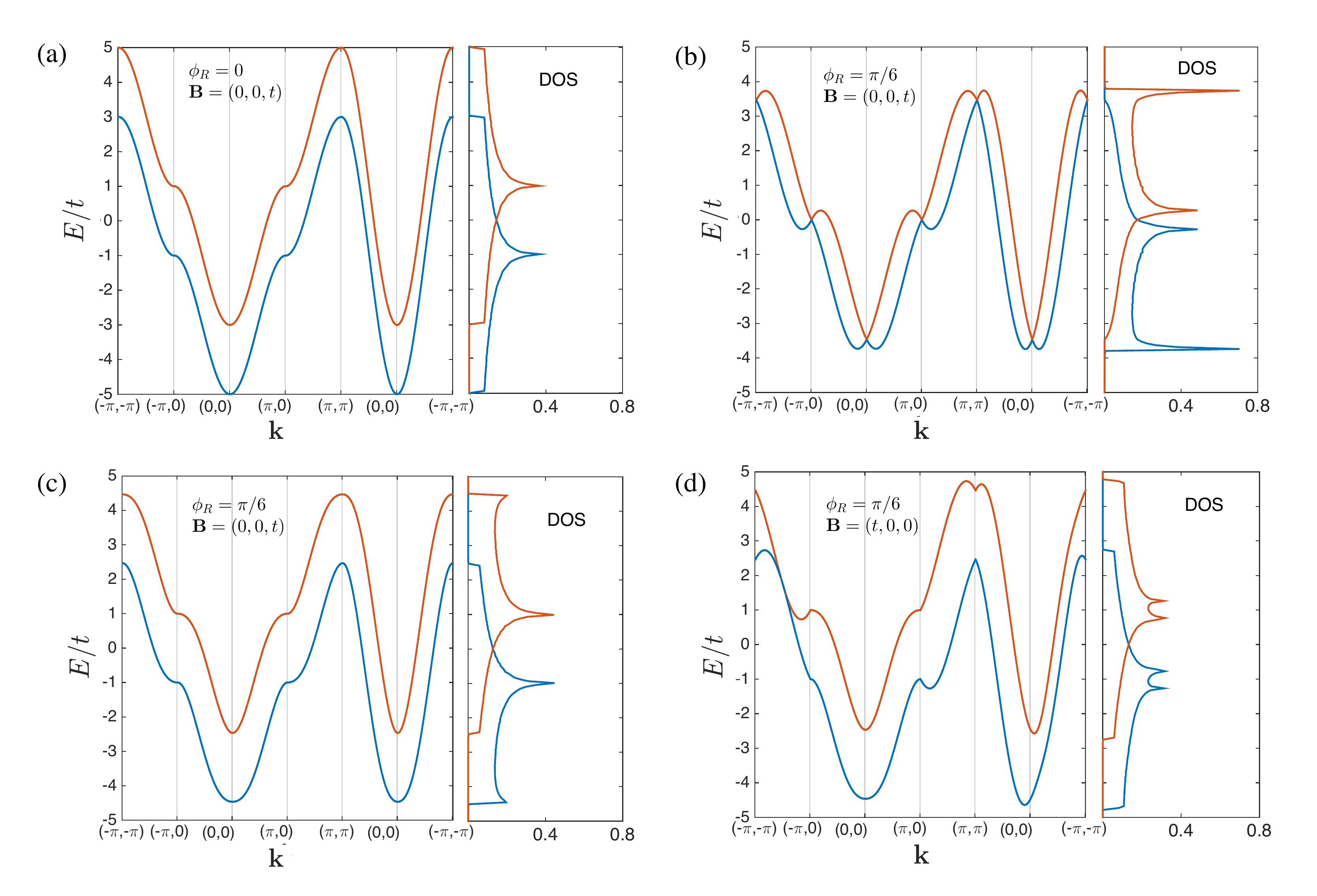}
  \caption{\label{fig:bands}
  Band dispersions given by Eq.~\eqref{eq:bands} and the respective densities of states for representative cases:
  (a) No Rashba splitting and finite exchange coupling to the background magnetization 
  leads to a constant vertical splitting of the bands.
  Parameters: $\PHIR=0$ ($t' = 2t$, $t'' = 0$), $J = t$.
  (b) Finite Rashba splitting and no background magnetization 
  leads to a $\VEC{k}$-dependent horizontal splitting of the bands.
  Parameters: $\PHIR=\pi/6$ ($t' = \sqrt{3}t$, $t'' = t$), $J = 0$.
  (c,d) When the Rashba splitting and the background magnetization are both finite, the dispersion depends on the orientation of the magnetization with respect to the lattice.
  Parameters: $\PHIR=\pi/6$ ($t' = \sqrt{3}t$, $t'' = t$), $J = t$.
  (c) When the magnetization is normal to the plane ($\VEC{B} \parallel \hat{\VEC{z}}$) the system has fourfold rotational symmetry.
  (d) When the magnetization is along a nearest-neighbor direction ($\VEC{B} \parallel \hat{\VEC{x}}$) the bands have a unidirectional shift in the perpendicular direction ($\hat{\VEC{y}}$).
  For these parameters we find two degeneracy 
  points, at $\VEC{k} = (0,-\pi/2)$ and $\VEC{k} = (\pm\pi,-\pi/2)$, one being visible in the figure.
  These degeneracies do not lead to any features in the DOS.}
\end{figure*}

We can immediately diagonalize the Hamiltonian,
\begin{equation}
  \MC{H}(\VEC{k}) = E_+(\VEC{k})\,P_+(\VEC{k}) + E_-(\VEC{k})\,P_-(\VEC{k}) \quad,
\end{equation}
where the band energies
\begin{equation}\label{eq:bands}
  E_\pm(\VEC{k}) = E_0(\VEC{k}) \mp \NORM{\VEC{b}(\VEC{k})} \quad,
\end{equation}
and the eigenvector projectors
\begin{equation}
   P_\pm(\VEC{k}) = \frac{1}{2}\left(\sigma^0 \pm \hat{\VEC{b}}(\VEC{k})\cdot\bm{\upsigma}\right)
  \quad,\qquad \hat{\VEC{b}}(\VEC{k}) = \frac{\VEC{b}(\VEC{k})}{\NORM{\VEC{b}(\VEC{k})}} \quad .
\end{equation}
The plus sign corresponds to the lower energy 
majority band and the minus sign to the higher energy minority band.
Band dispersions are plotted in Fig.~\ref{fig:bands} for some representative cases.
 
The electronic density of states (DOS) is given by
\begin{equation}\label{eq:dos}
  \rho(E) = \sum_{n=\pm}\int\!\frac{\ud\VEC{k}}{(2\pi)^2}\;\delta\big(E-E_{n}(\VEC{k})\big) \quad,
\end{equation}
which leads to the number of electrons per lattice site,
\begin{equation}\label{eq:elnumber}
  N_{\text{e}} = \int\!\frac{\ud\VEC{k}}{(2\pi)^2}\;\big(f_+(\VEC{k}) + f_-(\VEC{k})\big) = \int^{\EF}_{-\infty}\hspace{-1em}\ud E\;\rho(E) \quad.
\end{equation}
The integral in Eq.~\ref{eq:elnumber} is over the first Brillouin zone, and $f_n(\VEC{k}) = \Theta(\EF-E_{n}(\VEC{k}))$ is the occupation of the corresponding eigenstate $E_{n}(\VEC{k})$.
The coupling to the ferromagnetic background induces a net spin moment on the itinerant electrons, given by
\begin{equation}\label{eq:mspin}
  \VEC{M} = \int\!\frac{\ud\VEC{k}}{(2\pi)^2}\;\big(f_+(\VEC{k}) - f_-(\VEC{k})\big)\,\hat{\VEC{b}}(\VEC{k})
   = \int^{\EF}_{-\infty}\hspace{-1em}\ud E\;\VEC{m}(E) \quad,
\end{equation}
which defines the spin-polarized DOS (the net vector spin polarization at a given energy).
The energetics of the itinerant electrons can be obtained from the internal energy $U$.
At zero temperature,
\begin{equation}\label{eq:intenergy}
  U = \sum_{n=\pm}\int\!\frac{\ud\VEC{k}}{(2\pi)^2}\;f_n(\VEC{k})\,E_{n}(\VEC{k})
  = \int^{\EF}_{-\infty}\hspace{-1em}\ud E\;\rho(E)\,E \quad .
\end{equation}
Some properties of the internal energy are summarized in Appendix~\ref{app:statphys}.
Further insight can be gained by separating contributions to the internal energy into 
bare band, Rashba, and exchange contributions (cf.~Eq.~\eqref{eq:sdhamil}):
\begin{align}
   \label{eq:uintsplit}
   U &= \sum_{n=\pm}\int\!\frac{\ud\VEC{k}}{(2\pi)^2}\;f_n(\VEC{k})\,\Tr\,P_n(\VEC{k})\,\big(\MC{H}_0(\VEC{k}) + \MC{H}_{\text{R}}(\VEC{k}) + \MC{H}_{\text{B}}\big)\notag\\ 
   &= U_0 + U_{\text{R}} + U_{\text{B}} \quad .
\end{align}

For some calculations it is more convenient to employ the Green function
\begin{equation}\label{eq:gfspectral}
  G(\VEC{k},E) = \big(E- \MC{H}(\VEC{k})\big)^{-1} = \sum_{n=\pm}\frac{P_n(\VEC{k})}{E - E_n(\VEC{k})} \quad,
\end{equation}
which is related to the internal energy and its derivatives in Appendix~\ref{app:gfhf}.
For instance, the DOS is given by
\begin{equation}
  \rho(E) = -\frac{1}{\pi}\,\IM\,\Tr\!\int\!\frac{\ud\VEC{k}}{(2\pi)^2}\;G(\VEC{k},E) \quad ,
\end{equation}
and the spin-polarized DOS by
\begin{equation}\label{eq:mspindos}
  \VEC{m}(E) = -\frac{1}{\pi}\,\IM\,\Tr\!\int\!\frac{\ud\VEC{k}}{(2\pi)^2}\;\VEC{\upsigma}\;G(\VEC{k},E) \quad ,
\end{equation}
where the traces are over the spin components.

All ground state properties can be expressed in terms of the Green function, including correlation functions.
In particular, the static uniform spin susceptibility for a fixed number of electrons is given by (using Eqs.~\eqref{eq:sdhamil}, \eqref{eq:mspin} and \eqref{eq:mspindos}, and the property \eqref{eq:gfdiffparam})
\begin{align}\label{eq:susc0}
  \chi^{\alpha\beta} &= \left.\frac{\partial M^\alpha}{\partial B^\beta}\right|_{N_{\text{e}}}\notag\\
  &= \frac{1}{\pi}\,\IM\,\Tr\int^{\EF}_{-\infty}\hspace{-1em}\ud E\!\int\!\frac{\ud\VEC{k}}{(2\pi)^2}\;\sigma^\alpha\,G(\VEC{k},E)\,\sigma^\beta\,G(\VEC{k},E)\notag\\
  &\phantom{=}\;  -\frac{m^\alpha(\EF)\,m^\beta(\EF)}{\rho(\EF)} \quad.
\end{align}
The last term comes from ensuring that $\partial N_{\text{e}} / \partial B^\beta = 0$, as in the derivation of Eq.~\eqref{eq:eintdiff2gf}.
Its role is illustrated in Appendix~\ref{app:anatomynosoc} for a ferromagnetic system without SOC.
The susceptibility can also be expressed directly in terms of the eigenvalues and eigenvectors of the Hamiltonian, as summarized in Appendix~\ref{app:anatomy}.

Some comments on the numerical evaluation of the various quantities we 
consider are in order.
Every quantity is to be calculated at constant filling $N_{\text{e}}$, which requires an accurate determination of the Fermi energy $\EF$.
Keeping all other parameters fixed (magnetization orientation, etc.), $\EF$ is a monotonic function of $N_{\text{e}}$, so it can be efficiently determined using the bisection algorithm with high accuracy.
$\EF$ is iteratively refined until the computed $N_e$ is within a $\pm 10^{-8}$ range of the desired input value.
It follows from particle-hole symmetry that  $\EF = 0$ for $N_{\text{e}} = 1$.
The integrals over the Brillouin zone are done with a k-mesh of $1000\times 1000$ equidistant points.
To compute the DOS, the $\delta$-functions in Eq.~\eqref{eq:dos} are approximated by Lorentzian functions with a broadening $\eta = 10^{-3}\,t$.
All other quantities are computed by direct numerical summation of the contributions from each k-point, using either the analytical expressions or contour integration of the Green function expressions.
%%%%%%%%%%%%%%%%%%%%%%%%%%%%%%%%%%%%%%%%%%%%%%%%%%%%%%%%%%%%%%%%%%%%%%%%%%%%%%%%%%%%%%%%%%%%%%%%%%%%

%%%%%%%%%%%%%%%%%%%%%%%%%%%%%%%%%%%%%%%%%%%%%%%%%%%%%%%%%%%%%%%%%%%%%%%%%%%%%%%%%%%%%%%%%%%%%%%%%%%%
\section{Computing the magnetic anisotropy energy}\label{sec:mae}
%%%%%%%%%%%%%%%%%%%%%%%%%%%%%%%%%%%%%%%%%%%%%%%%%%%%%%%%%%%%%%%%%%%%%%%%%%%%%%%%%%%%%%%%%%%%%%%%%%%%
In our model, the MAE is due to the variation of the internal energy of the itinerant electrons
as the ferromagnetic background orientation rotates.
Following the arguments of Bloch and van Vleck~\cite{Bloch1931,vanVleck1937}, it is clear that the MAE vanishes
if there is no spin-orbit coupling, {\it i.e.} in our model if there is no Rashba coupling ($\PHIR = t'' = 0$).
Phenomenologically, the MAE is expanded in angular functions that respect the symmetry of the system.\cite{Skomski2008}
For the square lattice (effectively tetragonal symmetry),
\begin{equation}\label{eq:mae}
  U_{\text{MAE}}(\theta,\varphi) \approx K_2 \sin^2\theta + \left(K_4 + K_4'\cos4\varphi\right)\sin^4\theta \;,
\end{equation}
with $\theta$ and $\varphi$ the spherical angles describing the orientation of the ferromagnetic background.
It follows from perturbation theory arguments that
$K_{2n} \propto t''\, (t''/J)^{2n-1}$ with $n \geq 1$, as discussed for the present
model in Section~\ref{sec:halffilling}.
Higher-order anisotropy constants should decline rapidly in magnitude, 
as they are proportional to higher powers of the ratio between the 
spin-orbit interaction strength and the spin splitting, which is often small.

\begin{figure}[tb]
  \begin{tabular}{l}
    \includegraphics[width=0.4\textwidth]{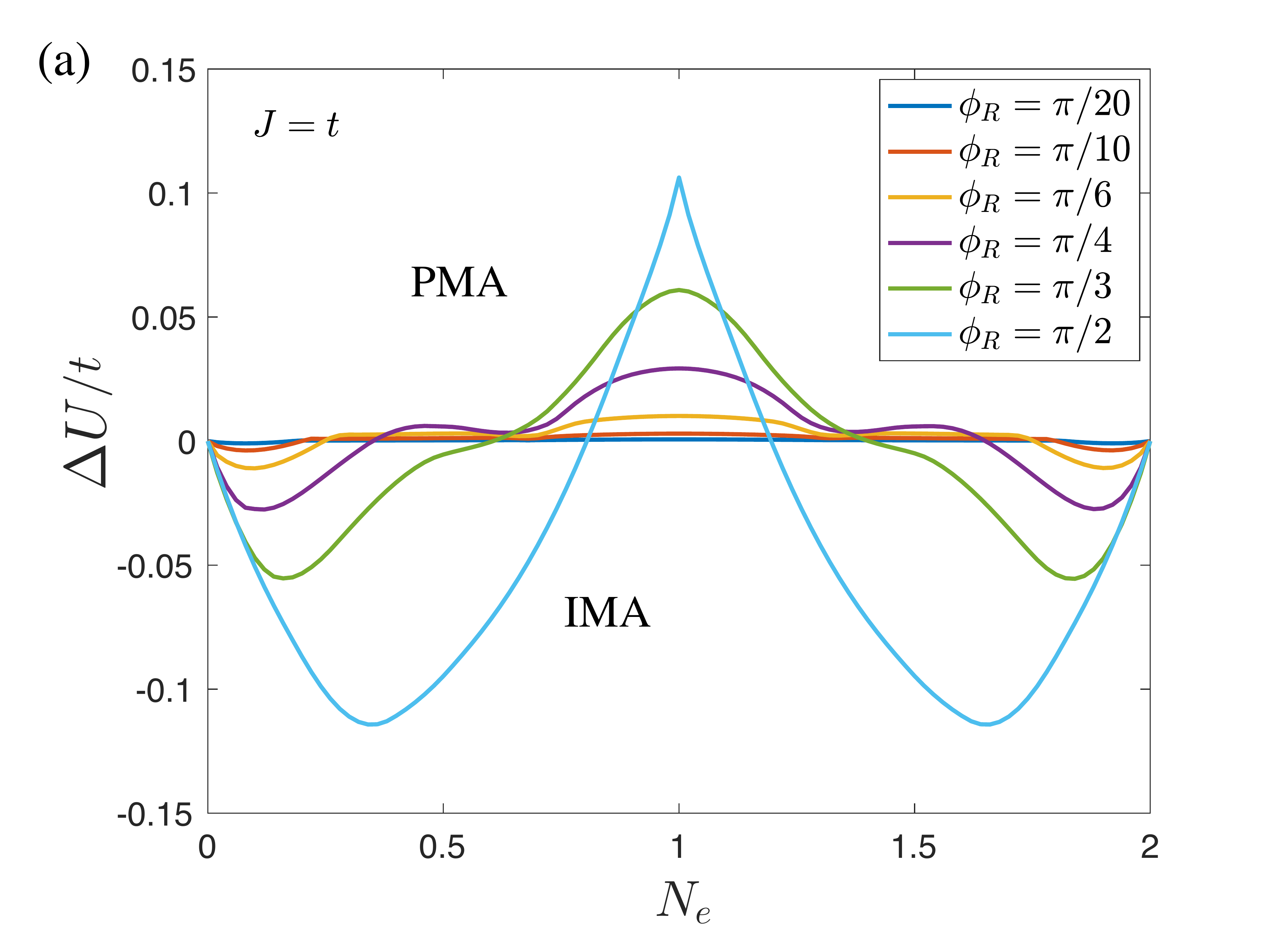} \\
    \includegraphics[width=0.4\textwidth]{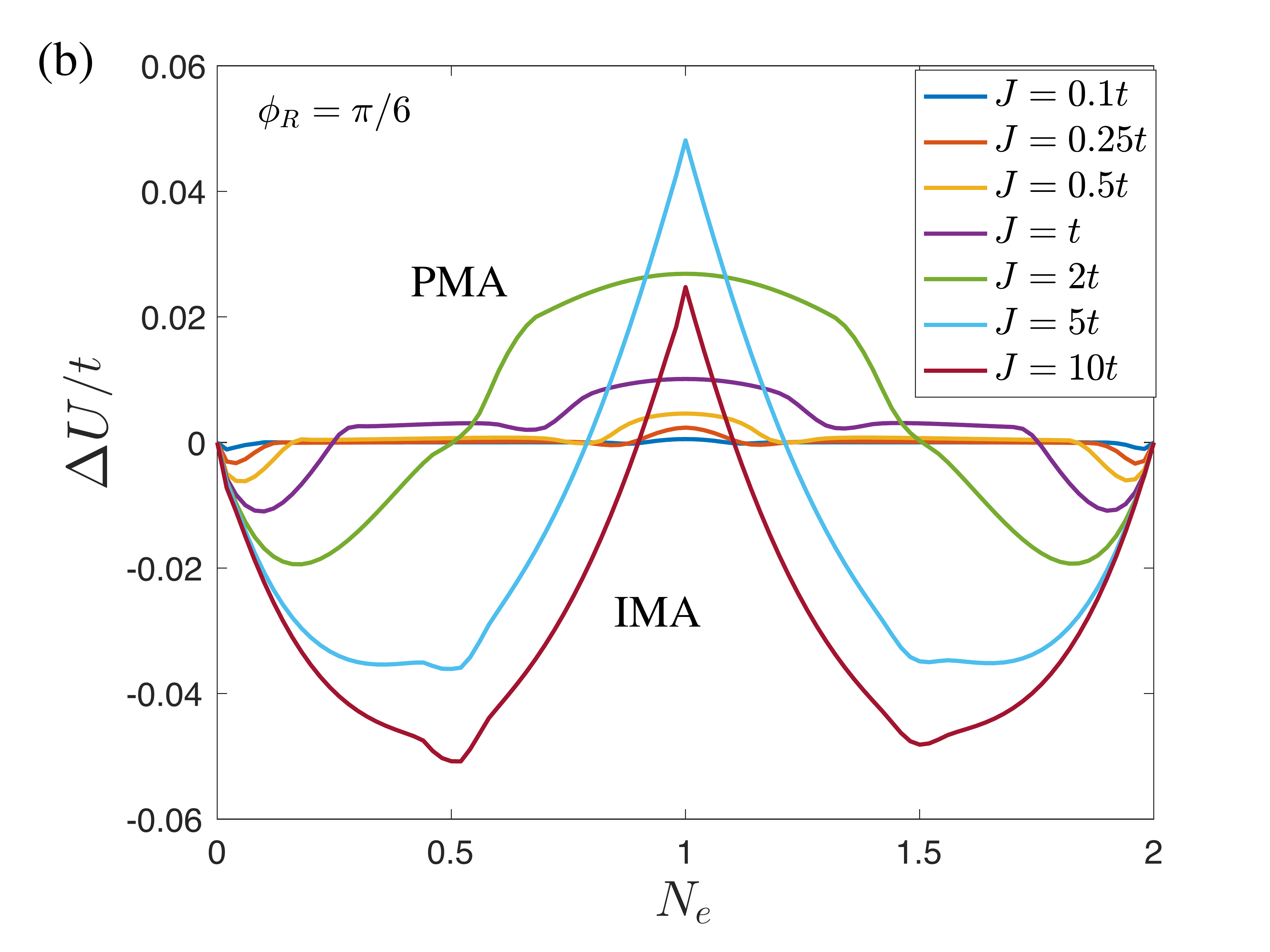}
  \end{tabular}
  \caption{\label{fig:maeglobal}
  Magnetic anisotropy energy $\Delta U = U(\nicefrac{\pi}{2},0) - U(0,0)$ {\it vs.} number of electrons per site.
  (a) For increasing Rashba strength and fixed coupling strength to the ferromagnetic background.
  Parameters: $J = t$.
  (b) For fixed Rashba strength and increasing coupling strength to the ferromagnetic background.
  Parameters: $\PHIR=\pi/6$ ($t' = \sqrt{3}t$, $t'' = t$).}
\end{figure}

The anisotropy constants can then be determined 
by fitting the angular dependence of the internal energy.
Keeping all other parameters fixed, the internal energy given by Eq.~\eqref{eq:intenergy} is an 
explicit function of the angles describing the ferromagnetic orientation, $U(\theta,\varphi)$.
Assuming that the model form in Eq.~\eqref{eq:mae} holds, evaluating the internal energy for three orientations is sufficient to fix the anisotropy.  The system will have PMA provided that both of the following inequalities are satisfied: 
\begin{equation}\label{eq:uintdiff}
  \left.\begin{array}{r}
  U(\nicefrac{\pi}{2},0) - U(0,0) = K_2 + K_4 + K_4' \\
  U(\nicefrac{\pi}{2},\nicefrac{\pi}{4}) - U(0,0) = K_2 + K_4 - K_4'
  \end{array}\right\} > 0\;.
\end{equation}
Often $K_4'$ can be neglected, and only two orientations of the magnetization need be considered.
In Fig.~\ref{fig:maeglobal} we show how the anisotropy energy goes from 
IMA $\rightarrow$ PMA $\rightarrow$ IMA as a function of the band filling, as already 
found in Ref.~\onlinecite{Kim2016a}.
In Sec.~\ref{sec:casestudies} we test the claim that the MAE is equal to  
half the magnetization direction dependence of the SOC energy.\cite{Laan1999,Antropov2014a}

We can also calculate the MAE in two alternative ways.
For a chosen orientation of the ferromagnetic background, we may compute either the magnetic 
torque (the first derivative of the internal energy with respect to the ferromagnetic moment orientation) 
or the curvature of the internal energy (the second derivative).
The Hellmann-Feynman theorem~\cite{Hellmann1937,Feynman1939} yields the first derivative
of the internal energy in a convenient form. 
(A detailed derivation is presented in Appendix~\ref{app:gfhf}.)
Using Eq.~\eqref{eq:hftheorem} we have
\begin{align}\label{eq:dudt}
  \frac{\partial U}{\partial\theta} &= -\VEC{M} \cdot \frac{\partial\VEC{B}}{\partial\theta} \nonumber\\
  &= \left(K_2 + 2\left(K_4 + K_4'\cos4\varphi\right)\sin^2\theta\right)\sin2\theta \quad, \\
  \label{eq:dudp}
  \frac{\partial U}{\partial\varphi} &= -\VEC{M} \cdot \frac{\partial\VEC{B}}{\partial\varphi} \nonumber\\
  &= -4\,K_4'\sin4\varphi\sin^4\theta \quad,
\end{align}
where $\VEC{M}$ is the spin magnetic moment of the electrons defined in Eq.~\ref{eq:mspin}, and $\VEC{B}$ is the effective magnetic field produced 
by the local moments defined in Eq.~\ref{eq:spinaxis}.
From the phenomenological expression for $U_{\text{MAE}}(\theta,\varphi)$, we see that the 
magnetic torque $\VEC{M}\times\VEC{B}$ vanishes for the high-symmetry
nearest- and next-nearest-neighbor directions 
($\theta=\pi/2$ and $\varphi=n\pi/4$, with $n\in\{0,1,\ldots,7\}$), and for magnetization normal to the
lattice plane ($\theta = 0,\pi$).

The second derivatives of the internal energy are particularly simple to evaluate for these high-symmetry directions, 
since cross derivatives involving both polar and azimuthal angles vanish.
We therefore only need to evaluate only $\partial^2 U/\partial\theta^2$ and $\partial^2 U/\partial\phi^2$.
Utilizing Eqs.~\eqref{eq:eintdiff2gf} and \eqref{eq:sdhamil}, we see that we require only the cartesian component of the spin susceptibility tensor for the plane perpendicular to a chosen magnetization direction (i.e.~we need only the transverse spin susceptibility).
For the high-symmetry directions the net spin moment of the itinerant electrons is aligned with the 
ferromagnetic background, $\VEC{M}\parallel\VEC{B}$, and so the second term 
in Eq.~\eqref{eq:susc0} vanishes for the transverse susceptibility.
For the in-plane high-symmetry directions, Eqs.~\eqref{eq:susc0} and \eqref{eq:eintdiff2gf} lead to
\begin{equation}\label{eq:d2udt2suscx}
  \frac{1}{2}\!\left.\frac{\partial^2 U}{\partial\theta^2}\right|_{\VEC{M} \parallel \hat{\VEC{x}}} = \frac{J^2}{2}\left(\frac{M}{J} - \chi^{zz}\right)
  = -K_2 - 2\left(K_4 + K_4'\right) \quad,
\end{equation}
\begin{equation}\label{eq:d2udp2suscx}
  \frac{1}{2}\!\left.\frac{\partial^2 U}{\partial\varphi^2}\right|_{\VEC{M} \parallel \hat{\VEC{x}}} = \frac{J^2}{2}\left(\frac{M}{J} - \chi^{yy}\right)
  = -8K_4' \quad,
\end{equation}
and for the polar magnetization orientation
\begin{equation}\label{eq:d2udt2suscz}
  \frac{1}{2}\!\left.\frac{\partial^2 U}{\partial\theta^2}\right|_{\VEC{M} \parallel \hat{\VEC{z}}} = \frac{J^2}{2}\left(\frac{M}{J} - \chi^{xx}\right)
  = K_2 \, \quad .
\end{equation}
The $M/J$ contribution comes from the first term on the right-hand side of \eqref{eq:eintdiff2gf}.

When $\VEC{M} \parallel \hat{\VEC{z}}$, the system has fourfold rotational symmetry from 
which it follows that $\chi^{xx} = \chi^{yy}$ and $\chi^{xy} = \chi^{yx} = 0$. 
We can gain further insight into the MAE by separating the transverse spin susceptibility into intraband and interband contributions, as explained in Appendix~\ref{app:anatomy}. To simplify this discussion, we subtract a common term
\begin{equation}\label{eq:chidrop}
  \bar{\chi} = \int\!\frac{\ud\VEC{k}}{(2\pi)^2}\,\frac{f_+(\VEC{k}) - f_-(\VEC{k})}{\NORM{\VEC{b}(\VEC{k})}} \quad,
\end{equation}
{from the quantities entering Eqs. \eqref{eq:d2udt2suscx} and \eqref{eq:d2udt2suscz}.
The $M/J$ term then becomes
\begin{equation}\label{eq:suscvol}
  \chi^0 = \frac{M}{J} - \bar{\chi} = \int\!\frac{\ud\VEC{k}}{(2\pi)^2}\,\frac{\VEC{B}\cdot\VEC{b}_{\text{R}}(\VEC{k})}{J^2}\,
  \frac{f_+(\VEC{k}) - f_-(\VEC{k})}{\NORM{\VEC{b}(\VEC{k})}} \quad,
\end{equation}
which we will refer to as the volume susceptibility.
The expression for the intraband part of the spin susceptibility follows from Eq.~\eqref{eq:suscintra} and does not contain $\bar{\chi}$}:
\begin{align}\label{eq:chiintra}
  \chi^{\alpha\alpha}_{\text{intra}} = \int\!\frac{\ud\VEC{k}}{(2\pi)^2}\;\big(\hat{b}_{\alpha}(\VEC{k})\big)^2 \sum_n \delta\big(\EF - E_n(\VEC{k})\big) \quad,
%  = \frac{\partial}{\partial\EF}\int\!\frac{\ud\VEC{k}}{(2\pi)^2}\;\big(\hat{b}_{\alpha}(\VEC{k})\big)^2\,\big(f_+(\VEC{k}) + f_-(\VEC{k})\big) \quad, \\
\end{align}
This term is however present in the interband part of the spin susceptibility, so we subtract it from Eq.~\eqref{eq:suscinter}:
\begin{align}
  \label{eq:chiinter}
  &\bar{\chi}^{\alpha\alpha}_{\text{inter}} = \chi^{\alpha\alpha}_{\text{inter}}-\bar{\chi}\notag\\ 
  &\hspace{0.85cm}=  -\!\int\!\frac{\ud\VEC{k}}{(2\pi)^2}\,\big(\hat{b}_{\alpha}(\VEC{k})\big)^2\,
  \frac{f_+(\VEC{k}) - f_-(\VEC{k})}{\NORM{\VEC{b}(\VEC{k})}} \quad.
\end{align}
In the previous two equations $\alpha = x,y,z$, and $\hat{b}_{\alpha}(\VEC{k})$ are the cartesian components of the unit vector defining the spin quantization axis for each $\VEC{k}$ (see Eq.~\eqref{eq:spinaxis}). %The importance of the above modified definitions of susceptibilities can be estabilished in the fact that they clearly separate interband and intraband contribution to the MAE, viz
We see that $\chi^{\alpha\alpha}_{\text{intra}}$ arises from the Fermi surface and is positive definite, 
while $\bar{\chi}^{\alpha\alpha}_{\text{inter}}$ arises from the Fermi sea and is negative definite.

Whether we have PMA or IMA can then be established in two ways.
When $\VEC{M} \parallel \hat{\VEC{x}}$ we have $\chi^{zz} = 0$, and the sign of the MAE is determined by $\chi^0$. 
On the other hand, for $\VEC{M} \parallel \hat{\VEC{z}}$ we find that $\chi^0 = 0$, so the sign of the MAE is decided by the competition between the intraband and interband contributions to the spin susceptibility $\chi^{xx}$.
The detailed analysis in Sec.~\ref{sec:casestudies} shows that both results are consistent, and can be given a meaningful interpretation.

Next we look more closely at the conditions that favor PMA.
Let $\VEC{M} \parallel \hat{\VEC{z}}$ and $|t''| \ll J$.
Making the constant matrix element approximation in Eq.~\eqref{eq:chiintra}, we find
\begin{align}
  \chi^{xx}_{\text{intra}} &\approx \left\langle \big(\hat{b}_x(\VEC{k})\big)^2\right\rangle \int\!\frac{\ud\VEC{k}}{(2\pi)^2}\sum_n \delta\big(\EF - E_n(\VEC{k})\big) \nonumber\\
  &= \frac{(t'')^2}{2J^2}\,\rho(\EF) \quad .
\end{align}
Here $\rho(\EF)$ is the total density of states at the Fermi energy.
The average of the matrix element was simplified by assuming that the exchange 
fields are much stronger than the spin-orbit fields, so that $\NORM{\VEC{b}(\VEC{k})} \approx J$.
Evaluating Eq.~\eqref{eq:chiinter} in the same way we obtain
\begin{equation}
  \bar{\chi}^{xx}_{\text{inter}} \approx -\frac{(t'')^2}{2J^3}\,M \quad .
\end{equation}
Combining these expressions we arrive at an appealing approximate form for the uniaxial anisotropy constant,
\begin{align}\label{eq:k2approx}
  K_2 &\approx K_{\text{ref}}\,\big(M - J \rho(\EF)\big) \quad,
\end{align}
where the scale of the anisotropy constant
\begin{equation}\label{eq:kref}
 K_{\text{ref}} = \frac{1}{4}\,\frac{(t'')^2}{J} \quad ,
\end{equation}
is a useful figure of merit for MAE.
As will be shown in Sec.~\ref{sec:halffilling}, this is the leading order contribution to 
$K_2$ for the gapped half-filled case ($M = 1$ and $\rho(\EF) = 0$).
We conclude that PMA is likely to be stable when 
the density-of-states at the Fermi level is small: Since $ 0 \leq M \leq 1$, we can expect PMA if $J \rho(\EF) \lesssim 1$.
The states at the Fermi level are the ones affected by SOC in the most important way, in energetic terms.
A large DOS at the Fermi level then translates to a large number of single-particle states states that gain the most energy from SOC once the magnetization is tilted away from the perpendicular direction, which explains why this contribution favors IMA. Eq.~\ref{eq:k2approx} is approximate but provides a useful reference point for the 
case studies discussed in detail in Sec.~\ref{sec:casestudies} below.

%%%%%%%%%%%%%%%%%%%%%%%%%%%%%%%%%%%%%%%%%%%%%%%%%%%%%%%%%%%%%%%%%%%%%%%%%%%%%%%%%%%%%%%%%%%%%%%%%%%%
\section{Perturbation theory for the gapped half-filled case}\label{sec:halffilling}
%%%%%%%%%%%%%%%%%%%%%%%%%%%%%%%%%%%%%%%%%%%%%%%%%%%%%%%%%%%%%%%%%%%%%%%%%%%%%%%%%%%%%%%%%%%%%%%%%%%%
The simplest limit to consider is the case in which the ferromagnetic exchange splitting is 
large enough to produce a gap.  In the half-filled ferromagnetic insulator case,
the Fermi level lies in this gap, the majority band is full, 
$f_+(\VEC{k}) = 1$, and the minority band is empty, $f_-(\VEC{k}) = 0$.
The ferromagnetic insulator was found numerically to have PMA, both in our 
calculations and in Ref.~\onlinecite{Kim2016a}. 
Now we shall prove this property analytically.
Starting from Eqs.~\eqref{eq:spinaxis}, \eqref{eq:bands} and \eqref{eq:intenergy}, the internal 
energy for this case is simply
\begin{align}
  U = \int\!\frac{\ud\VEC{k}}{(2\pi)^2}\;\big(E_0(\VEC{k}) - \NORM{\VEC{b}(\VEC{k})}\big) \quad .
\end{align}
We see that only the second term in the integrand contains information about the orientation of the ferromagnetic background, given by the angles $\theta$ and $\varphi$.

We next expand the spin splitting $\NORM{\VEC{b}(\VEC{k})}$ in order to extract the $\theta$ and $\varphi$ dependence:
\begin{align}
  \NORM{\VEC{b}(\VEC{k})} &= \sqrt{\NORM{\VEC{b}_{\text{R}}(\VEC{k})}^2 + J^2 + 2\,\VEC{B}\cdot\VEC{b}_{\text{R}}(\VEC{k})} \notag\\
   &= b_0(\VEC{k})\sqrt{1 + \cos\gamma(\VEC{k})} \notag\\
   &= b_0(\VEC{k}) \sum_{n=0}^{\infty} \binom{\frac{1}{2}}{n} \big(\!\cos\gamma(\VEC{k})\big)^n \quad .
\end{align}
Here 
\begin{subequations}
\begin{align}
  b_0(\VEC{k}) &= \sqrt{\NORM{\VEC{b}_{\text{R}}(\VEC{k})}^2 + J^2} \quad, \\
  \cos\gamma(\VEC{k}) &= \frac{2\,\VEC{B}\cdot\VEC{b}_{\text{R}}(\VEC{k})}{\NORM{\VEC{b}_{\text{R}}(\VEC{k})}^2 + J^2} \quad .
\end{align}
\end{subequations}
The expansion can be written more explicitly in the form  
\begin{widetext}
\begin{align}
  \NORM{\VEC{b}(\VEC{k})} &= \sum_{n=0}^{\infty} \binom{\frac{1}{2}}{n} \frac{\left(2 J\,t''\right)^n}{\left(b_0(\VEC{k})\right)^{2n-1}}
  \left(\sin\theta\right)^n \left(\sin k_y\cos\varphi - \sin k_x\sin\varphi\right)^n \notag\\
  &= \sum_{n=0}^{\infty} \binom{\frac{1}{2}}{n} \frac{\left(2 J\,t''\right)^n}{\left(b_0(\VEC{k})\right)^{2n-1}} \left(\sin\theta\right)^n
  \sum_{p=0}^n \binom{n}{p} (-1)^p \left(\sin k_y\cos\varphi\right)^p \left(\sin k_x\sin\varphi\right)^{n-p} \notag\\
  &= \sum_{n=0}^{\infty} \sum_{p=0}^n B_n^p(\VEC{k}) \left(\sin\theta\right)^n \left(\cos\varphi\right)^p \left(\sin\varphi\right)^{n-p} \quad ,
\end{align}
\end{widetext}
with expansion coefficients
\begin{equation}
  B_n^p(\VEC{k}) = (-1)^p\,2^n\,\binom{\frac{1}{2}}{n} \binom{n}{p}\,\frac{\left( J\,t''\right)^n\left(\sin k_y\right)^p \left(\sin k_x\right)^{n-p}}{\big(\NORM{\VEC{b}_{\text{R}}(\VEC{k})}^2 + J^2\big)^{n-\frac{1}{2}}} \quad .
\end{equation}

The internal energy then has the corresponding expansion
\begin{equation}
  U(\theta,\varphi) = \sum_{n=0}^{\infty} \sum_{p=0}^n U_n^p \left(\sin\theta\right)^n \left(\cos\varphi\right)^p \left(\sin\varphi\right)^{n-p} \quad,
\end{equation}
with the coefficients
\begin{equation}
  U_n^p = -\!\int\!\frac{\ud\VEC{k}}{(2\pi)^2}\;B_n^p(\VEC{k}) \quad .
\end{equation}
Because the integrand is odd under $k_x \rightarrow -k_x$ and $k_y \rightarrow -k_y$ 
$U_{n}^{2p+1} = 0$ and $U_{2n+1}^{k} = 0$, {\it i.e.} only terms even in both $p$ and $n$ survive.
In combination with the symmetry of the binomial coefficients, we also 
have $U_{2n}^{2n-2p} = U_{2n}^{2p}$.
It follows that the first terms in the expansion are
\begin{align}\label{eq:maeexpansion}
  &U(\theta,\varphi) \approx U_0^0 + U_2^0 \sin^2\theta \notag\\
  &+ \left(\frac{6 U_4^0 + U_4^2}{8} + \frac{2 U_4^0 - U_4^2}{8}\cos4\varphi\right) \sin^4\theta  \quad,
\end{align}
in agreement with the phenomenological form given in Eq.~\eqref{eq:mae}.

\begin{figure}[tb]
  \begin{tabular}{c}
  \includegraphics[width=0.4\textwidth]{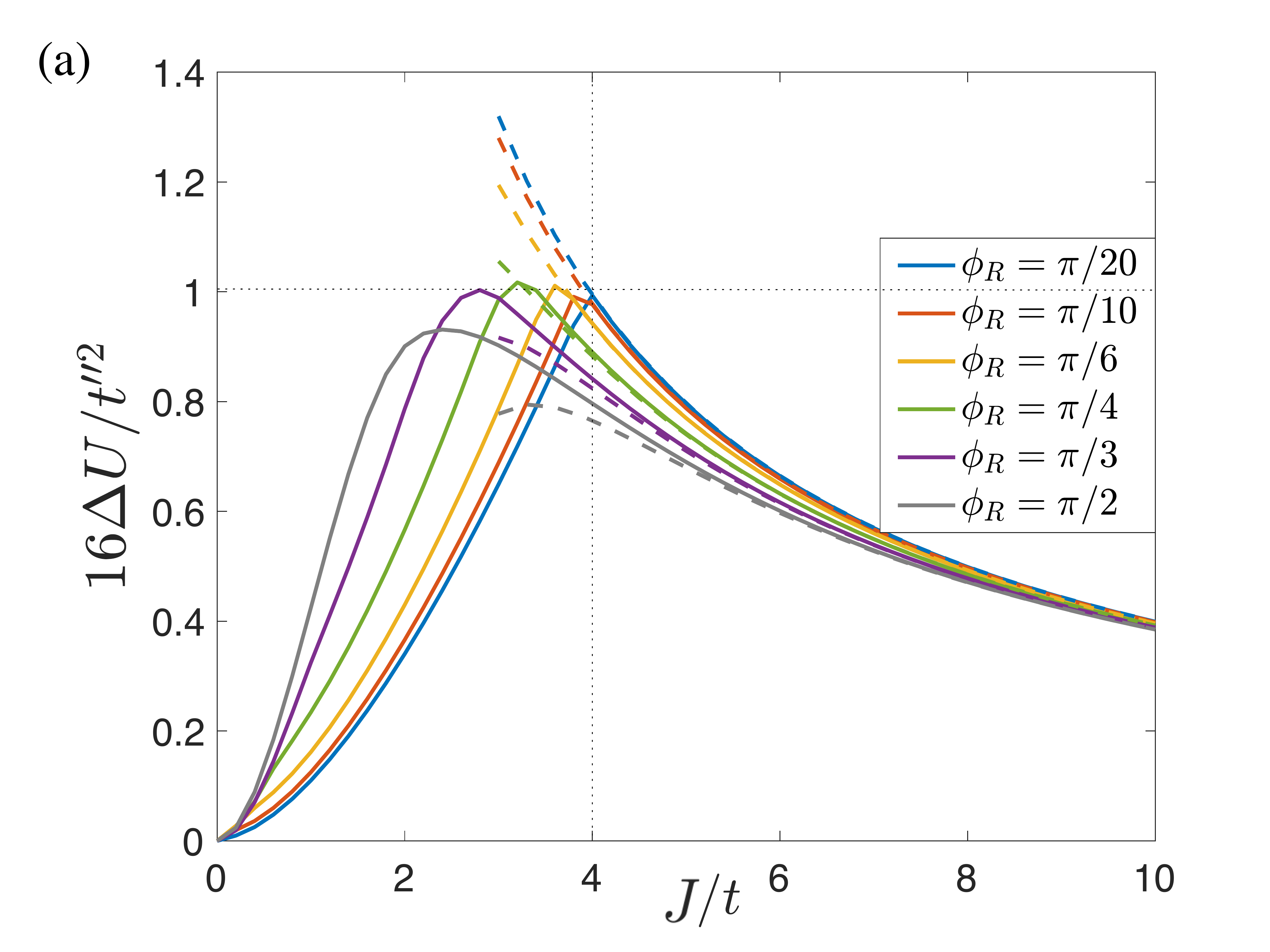} \\
  \includegraphics[width=0.4\textwidth]{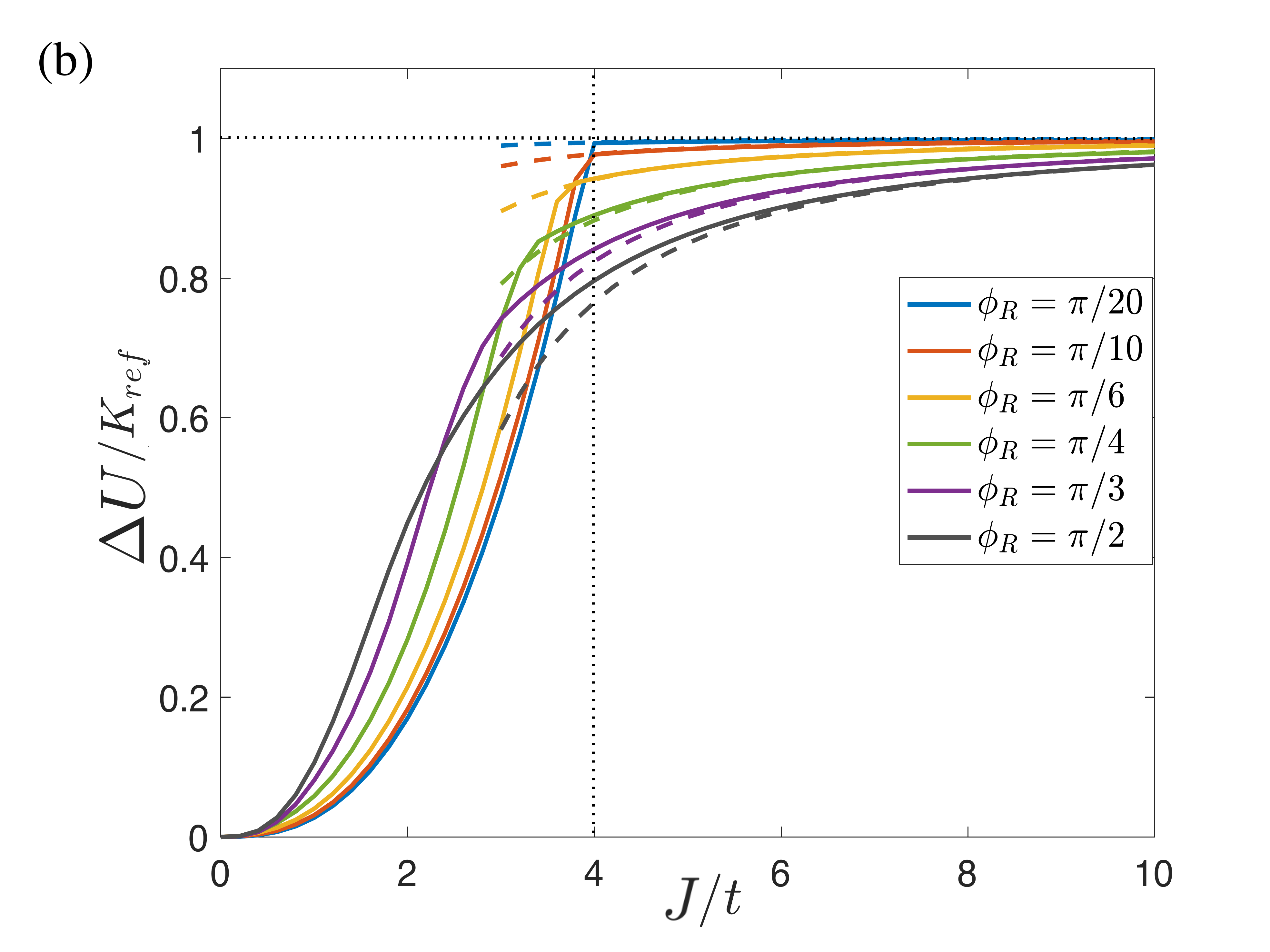}
  \end{tabular}
  \caption{\label{fig:mae_pt}
  MAE from perturbation theory for the half-filled case, $N_{\text{e}} = 1$, as a function the  
  coupling strength to the ferromagnetic background $J$ (in units of the hopping strength $t$).
  The solid lines are the numerically calculated internal energy 
   differences $\Delta U = U(\nicefrac{\pi}{2},0) - U(0,0)$.
  The dashed lines are the corresponding combination of anisotropy 
  coefficients in Eq.~\eqref{eq:uintdiff}, using the analytical forms of Eq.~\eqref{eq:mae_pt}.
  The vertical dotted line marks the closing of the gap in the weak SOC limit.
  We plot the results in two ways.
  (a) The energy axis is scaled by $t''^2/16$, to factor out the expected dependence on SOC strength.
  For a given Rashba interaction strength, 
  anisotropy energy is largest when the exchange coupling is just strong enough 
  to open a gap.
  (b) Anisotropy energy in units of $K_{\text{ref}} = (t'')^2/4J$, the form approached in the large $J$ limit.}
\end{figure}

For the gapped half-filled case, it is consistent to expand 
the integrand in the $\NORM{t''} \ll J$ limit.  
\begin{subequations}
   \begin{equation}
      B_0^0(\VEC{k}) \approx J  + \frac{\NORM{\VEC{b}_{\text{R}}(\VEC{k})}^2}{2J} \left(1 - \frac{\NORM{\VEC{b}_{\text{R}}(\VEC{k})}^2}{4J^2}\right) \quad ,
   \end{equation}
   \begin{equation}
      B_2^0(\VEC{k}) \approx -\frac{(t'')^2}{2J} \sin^2 k_x \left(1 - \frac{3\NORM{\VEC{b}_{\text{R}}(\VEC{k})}^2}{2J^2}\right) \quad ,
   \end{equation}
   \begin{equation}
      B_4^0(\VEC{k}) \approx -\frac{5{t''}^4}{8J^3} \sin^4 k_x \quad,
   \end{equation}
   \begin{equation}
      B_4^2(\VEC{k}) \approx -\frac{15(t'')^4}{4J^3} \sin^2 k_x \sin^2 k_y \quad .
   \end{equation}
\end{subequations}

The following integral can then be used to generate all $U_n^p$ coefficients:
\begin{align}\label{eq:genpoly}
  &\MC{I}_\ell(x,y) = \frac{1}{(2\pi)^2}\!\int_{-\pi}^{\pi}\hspace{-0.5em}\ud k_x\!\int_{-\pi}^{\pi}\hspace{-0.5em}\ud k_y\,\left(x\sin^2 k_x + y\sin^2 k_y\right)^\ell \notag\\
  &= \sum_{k=0}^\ell \binom{\ell}{k}\,x^k\,y^{\ell-k}\,\frac{4}{\pi^2}\!\int_{0}^{\frac{\pi}{2}}\hspace{-0.5em}\ud k_x\,\left(\sin k_x\right)^{2k}\notag\\
  &\hspace{4cm}\times\!\int_{0}^{\frac{\pi}{2}}\hspace{-0.5em}\ud k_y\,\left(\sin k_y\right)^{2(\ell-k)} \notag\\
  &= \sum_{k=0}^\ell \binom{\ell}{k} \frac{\big(2k-1\big)!!}{\big(2k\big)!!}\,\frac{\big(2(\ell-k)-1\big)!!}{\big(2(\ell-k)\big)!!}\;x^k\,y^{\ell-k} \quad.
\end{align}

For the general case of the integrand we derive
\begin{widetext}
\begin{equation}\label{eq:genpolyd}
  \frac{(\ell\!-\!m\!-\!n)!}{\ell!}\,\frac{\partial^{m+n}\MC{I}_\ell}{\partial x^m \partial y^n}(1,1)
  = \int\!\frac{\ud\VEC{k}}{(2\pi)^2}\;\big(\!\sin^2 k_x\big)^m\,\big(\!\sin^2 k_y\big)^n \left(\sin^2 k_x + \sin^2 k_y\right)^{\ell-m-n} \quad .
\end{equation}
\end{widetext}
The polynomials that will be needed in the following are
\begin{align}
  &\MC{I}_0(x,y) = 1, \quad \MC{I}_1(x,y) = \frac{x + y}{2},\notag\\
  & \MC{I}_2(x,y) = \frac{3}{8}\left(x^2+y^2\right) + \frac{1}{2}\,xy \quad.
\end{align}

The coefficients in the expansion of the internal energy are (skipping the constant shift of the energy)
\begin{subequations}\label{eq:mae_pt}
\begin{align}
  & U_2^0 = \frac{1}{2}\,\frac{(t'')^2}{J}\,\frac{\partial I_1}{\partial x}(1,1) - \frac{3}{8}\,\frac{(t'')^4}{J^3}\,\frac{\partial I_2}{\partial x}(1,1)\notag\\ 
  &= \frac{1}{4}\,\frac{(t'')^2}{J} - \frac{15}{64}\,\frac{(t'')^4}{J^3} \quad ,
\end{align}
\begin{equation}
  U_4^0 = \frac{5}{16}\,\frac{(t'')^4}{J^3}\,\frac{\partial^2 I_2}{\partial x^2}(1,1)
  = \frac{15}{64}\,\frac{(t'')^4}{J^3} \quad ,
\end{equation}
\begin{equation}
  U_4^2 = \frac{15}{8}\,\frac{(t'')^4}{J^3}\,\frac{\partial^2 I_2}{\partial x \partial y}(1,1)
  = \frac{15}{16}\,\frac{(t'')^4}{J^3} \quad .
\end{equation}
\end{subequations}

From Eqs.~\eqref{eq:mae} and \eqref{eq:maeexpansion}, the anisotropy coefficients are then
\begin{align}
  &K_2 = \frac{1}{4}\,\frac{(t'')^2}{J} - \frac{15}{32}\,\frac{(t'')^4}{J^3}, \notag\\
  &K_4 = \frac{75}{256}\,\frac{(t'')^4}{J^3} \quad,\qquad K_4' = -\frac{K_4}{5} \quad.
\end{align}
This proves that the gapped half-filled case always displays PMA (when perturbation theory is valid).
The fourth order correction to $K_2$ weakens the anisotropy, but $K_4$ reinforces its easy-axis character.
The in-plane anisotropy is weak when compared to the uniaxial one, and favors alignment along the 
nearest-neighbor directions.
Appendix~\ref{app:anatomywithsoc} derives the same results starting 
from the transverse spin susceptibility.

Fig.~\ref{fig:mae_pt} shows the region of validity and the breakdown of perturbation theory for this case.
The maximum value of the PMA is obtained when the gap between the bands is about to close (e.g.~when 
$J \approx 4t$ for small $\phi_{\text{R}}$ or $t'' \ll t'$), which sets a limit on 
how much the PMA can be enhanced by reducing the magnitude of $J$.

%%%%%%%%%%%%%%%%%%%%%%%%%%%%%%%%%%%%%%%%%%%%%%%%%%%%%%%%%%%%%%%%%%%%%%%%%%%%%%%%%%%%%%%%%%%%%%%%%%%%
\section{Three case studies}\label{sec:casestudies}
%%%%%%%%%%%%%%%%%%%%%%%%%%%%%%%%%%%%%%%%%%%%%%%%%%%%%%%%%%%%%%%%%%%%%%%%%%%%%%%%%%%%%%%%%%%%%%%%%%%%
We now present a detailed analysis of the MAE for three different choices of model parameters, meant to illustrate different physical regimes at the ferromagnet/heavy-metal interface: (i) strong exchange ($J \gg t' \gg t''$), (ii) intermediate exchange ($J \sim t' \gg t''$), and (iii) weak exchange ($t' \gg J \sim t''$).
We fix the SOC strength to be smaller than the non-relativistic bandwidth, by setting $\phi_{\text{R}} = \pi/20$ ($t' = 2.0t$ and $t'' = 0.3t$).
The three case studies are then defined by how the exchange energy due to the ferromagnetic coupling 
compares to these two energy scales.
We shall compare the local characterization of the MAE via the susceptibility with the global characterization 
via internal energy differences.
For the present model, the contribution to the MAE from the volume susceptibility 
(Eq.~\eqref{eq:suscvol}) vanishes when $\VEC{M} \parallel \hat{\VEC{z}}$, while it is the 
only non-vanishing contribution for $\VEC{M} \parallel \hat{\VEC{x}}$.

\begin{figure*}[tb]
  \includegraphics[width=1.0\textwidth]{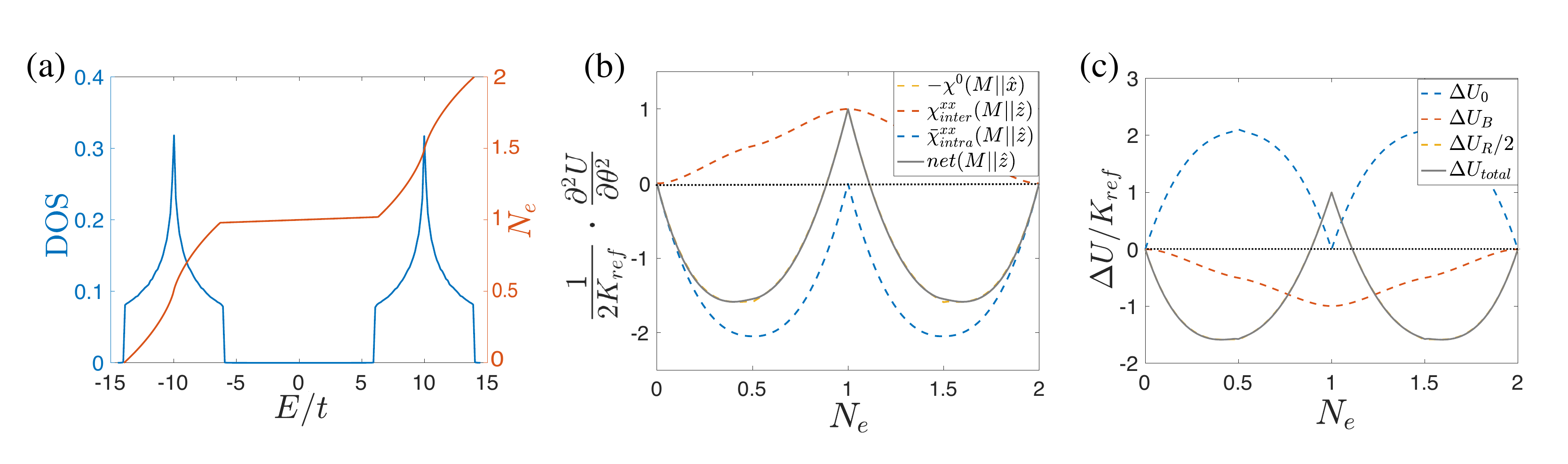}
  \caption{\label{fig:maecase1}
  MAE for the strong exchange case, $J \gg t' \gg t''$.
  (a) Total DOS and number of electrons as a function of energy, for $\VEC{M} \parallel \hat{\VEC{z}}$. 
  (b) MAE from the second derivatives of the band energy, from the connection to its phenomenological form.
  For $\VEC{M} \parallel \hat{\VEC{x}}$ (Eq.~\ref{eq:d2udt2suscx}), only $\chi^0$ contributes (Eq.~\eqref{eq:suscvol}).
  For $\VEC{M} \parallel \hat{\VEC{z}}$ (Eq.~\ref{eq:d2udt2suscz}), $\chi^0$ does not contribute, and we plot the intraband (Eq.~\eqref{eq:chiintra}) and interband (Eq.~\eqref{eq:chiinter}) contributions from the uniform spin susceptibility, as well as the net result.
  (c) Internal energy differences $\Delta U = U(\nicefrac{\pi}{2},0) - U(0,0)$, decomposed using Eq.~\ref{eq:uintsplit}.
  The curve showing half of the difference in the SOC energy overlaps almost perfectly with the net internal energy differences, which in turn agrees very well with the results obtained from the susceptibility calculations, c.f.~panel (b).
  Parameters: $J = 10t$ and $\PHIR=\pi/20$ ($t' = 2.0t$, $t'' = 0.3t$).
  }
\end{figure*}

We first consider the case where the exchange energy dominates, by setting $J = 10 t$.
This leads to two well-separated bands, as shown in Fig.~\ref{fig:maecase1}(a).
Fig.~\ref{fig:maecase1}(b) estimates the MAE from the spin susceptibility, for two stable orientations of the ferromagnetic background.
We see that for most values of $N_{\text{e}}$ we find IMA, with PMA only in a narrow range around $N_{\text{e}} = 1$.
When $\VEC{M} \parallel \hat{\VEC{z}}$ (Eq.~\ref{eq:d2udt2suscz}), the interband contribution to the susceptibility (Eq.~\eqref{eq:chiinter}) favors PMA, while the intraband contribution (Eq.~\eqref{eq:chiintra}) favors IMA.
The amplitude of the intraband contribution is larger than the interband one, 
and is maximized when the Fermi level is at the Van Hove singularity in the DOS of each band.
When $N_{\text{e}} = 1$ and $\VEC{M} \parallel \hat{\VEC{z}}$, the intraband contribution must vanish 
because the system is gapped.  
Only the interband term remains finite and it favors PMA.
When $\VEC{M} \parallel \hat{\VEC{x}}$ (Eq.~\ref{eq:d2udt2suscx}), the volume susceptibility (Eq.~\eqref{eq:suscvol}) is the only non-zero contribution, and reproduces essentially the same MAE as found 
for $\VEC{M} \parallel \hat{\VEC{z}}$.
This agreement shows that the higher-order anisotropy constants ($K_4$ and $K_4'$) are very small when compared with $K_2$, as anticipated from perturbation theory.
Fig.~\ref{fig:maecase1}(c) plots the MAE from the band energy difference between $\VEC{M} \parallel \hat{\VEC{x}}$ and $\VEC{M} \parallel \hat{\VEC{z}}$.
The MAE from this approach is in perfect agreement with the one extracted from the susceptibility.

Decomposing the band energy into its constituents (see Eq.~\eqref{eq:uintsplit}) we see that: (i) the anisotropy of the non-relativistic kinetic energy ($\Delta U_0$) matches the intraband contribution to the susceptibility ($\VEC{M} \parallel \hat{\VEC{z}}$), (ii) the anisotropy of the spin polarization 
energy ($\Delta U_{\text{B}}$) matches the interband contribution to the susceptibility ($\VEC{M} \parallel \hat{\VEC{z}}$), and (iii) half of the anisotropy of the Rashba energy ($\Delta U_{\text{R}}/2$) matches the contribution from the 
volume susceptibility ($\VEC{M} \parallel \hat{\VEC{x}}$).
We have verified the observation by van der Laan~\cite{Laan1999} and Antropov~\cite{Antropov2014a},
that the MAE is close to half of the anisotropy in the SOC (Rashba) energy, 
as predicted when SOC is treated as a weak perturbation. 

The behavior of the MAE can be qualitatively explained by the approximate formula in Eq.~\eqref{eq:kref}.
We find PMA near half-filling, as expected.
Moving from electron per site $N_{\text{e}} = 1$ to $N_{\text{e}} = 0$, the interband contribution is 
accurately proportional to 
$M$, which decreases monotonically to zero.
The intraband contribution qualitatively follows $\rho(\EF)$, which increases up to the Van Hove singularity and then decreases again, but the functional forms are not identical.
The intraband contribution is thus more sensitive to the constant matrix element approximation made in deriving Eq.~\eqref{eq:kref} than the interband contribution.
The transition from PMA to IMA is predicted by the $M \approx J\rho(\EF)$ criterion of Eq.~\eqref{eq:kref}
to occur at $N_{\text{e}} = 0.9$, in good agreement with the exact results.

\begin{figure*}[tb]
  \includegraphics[width=1.0\textwidth]{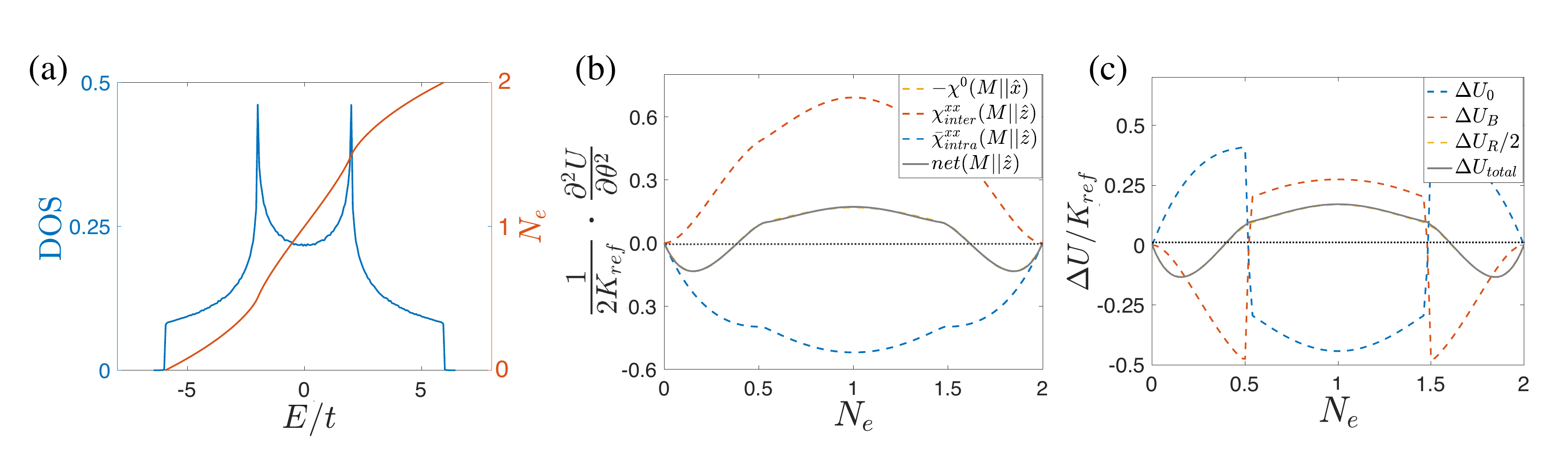}
  \caption{\label{fig:maecase2}
  MAE for the intermediate exchange case, $J \sim t' \gg t''$.
  (a) Total DOS and number of electrons as a function of Fermi energy, for $\VEC{M} \parallel \hat{\VEC{z}}$.
  (b) MAE from the second derivatives of the band energy, from the connection to its phenomenological form.
  For $\VEC{M} \parallel \hat{\VEC{x}}$ (Eq.~\ref{eq:d2udt2suscx}), only $\chi^0$ contributes (Eq.~\eqref{eq:suscvol}).
  For $\VEC{M} \parallel \hat{\VEC{z}}$ (Eq.~\ref{eq:d2udt2suscz}), $\chi^0$ does not contribute, and we plot the intraband (Eq.~\eqref{eq:chiintra}) and interband (Eq.~\eqref{eq:chiinter}) contributions from the uniform spin susceptibility, as well as the net result. 
  (c) Internal energy differences $\Delta U = U(\nicefrac{\pi}{2},0) - U(0,0)$, decomposed using Eq.~\ref{eq:uintsplit}.
  The curve showing half of the difference in the SOC energy overlaps almost perfectly with the net internal energy differences, which in turn agrees very well with the results obtained from the susceptibility calculations, c.f.~panel (b).
  Parameters: $J = t'$ and $\PHIR=\pi/20$ ($t' = 2.0t$, $t'' = 0.3t$).
  }
\end{figure*}

Next we consider the case where the exchange energy is comparable to the non-relativistic bandwidth, by setting $J = t'$.
Now the two bands overlap, as shown in Fig.~\ref{fig:maecase2}(a), with minority band occupation beginning for $N_{\text{e}} > 0.5$ ($\EF > -2t$), and the lower band being completely full for $N_{\text{e}} > 1.5$ ($\EF > 2t$).
This intermediate exchange coupling strength case is applicable to many ferromagnetic metals.  
Fig.~\ref{fig:maecase2}(b) estimates the MAE from the spin susceptibility, and shows that PMA is found in a much wider range of $N_{\text{e}}$ than in the strong exchange interaction case.
This was expected from Eq.~\eqref{eq:k2approx} when comparing to the previous case, as now $J$ is ten times weaker, so the condition $M \approx J\rho(\EF)$ is satisfied for a smaller value of $N_{\text{e}}$.
Comparing Eq.~\eqref{eq:chiintra} and Eq.~\eqref{eq:chiinter}, it appears that the Fermi sea term can be enhanced by reducing the k-dependent spin splitting $\NORM{\VEC{b}(\VEC{k})}$, which we achieved by weakening $J$, so that now the interband contribution has a larger amplitude than the intraband one.  However, near $N_{\text{e}} = 0$ (likewise near $N_{\text{e}} = 2$), the intraband contribution is linear in
 $N_{\text{e}}$ while the interband one is quadratic, so that the former can overtake the latter, and thus favors IMA.
As already shown in Fig.~\ref{fig:mae_pt}, the MAE reaches only 20\% of $K_{\text{ref}}$ at $N_{\text{e}} = 1$ (gapless system), in line with the discussion of Sec.~\ref{sec:halffilling}.
Fig.~\ref{fig:maecase2}(c) plots the MAE from the band energy difference between 
$\VEC{M} \parallel \hat{\VEC{x}}$ and $\VEC{M} \parallel \hat{\VEC{z}}$ and its decomposition.
Once again the band energy difference agrees very well with the results obtained from the susceptibility calculations, and with the estimate of $\Delta U_{\text{R}}/2$.
The previous identifications between the intraband and interband contributions to the susceptibility and the anisotropies of the non-relativistic kinetic energy and of the spin polarization energy, respectively, are seen to hold only while one of the bands is either completely empty ($N_{\text{e}} < 0.5$) or completely full ($N_{\text{e}} > 1.5$).
Although those two contributions to the energy exhibit discontinous behavior when both bands become partially filled, their sum is continuous, as can be concluded from $\Delta U_{\text{total}}$.  This shows that the energetic competition between the Rashba SOC and the coupling to the ferromagnetic background is settled differently when either only one or when both bands are partially filled, presumably due to an allowed transfer of electronic occupation between the two bands at the Fermi energy in the latter case.

\begin{figure*}[tb]
  \includegraphics[width=1.0\textwidth]{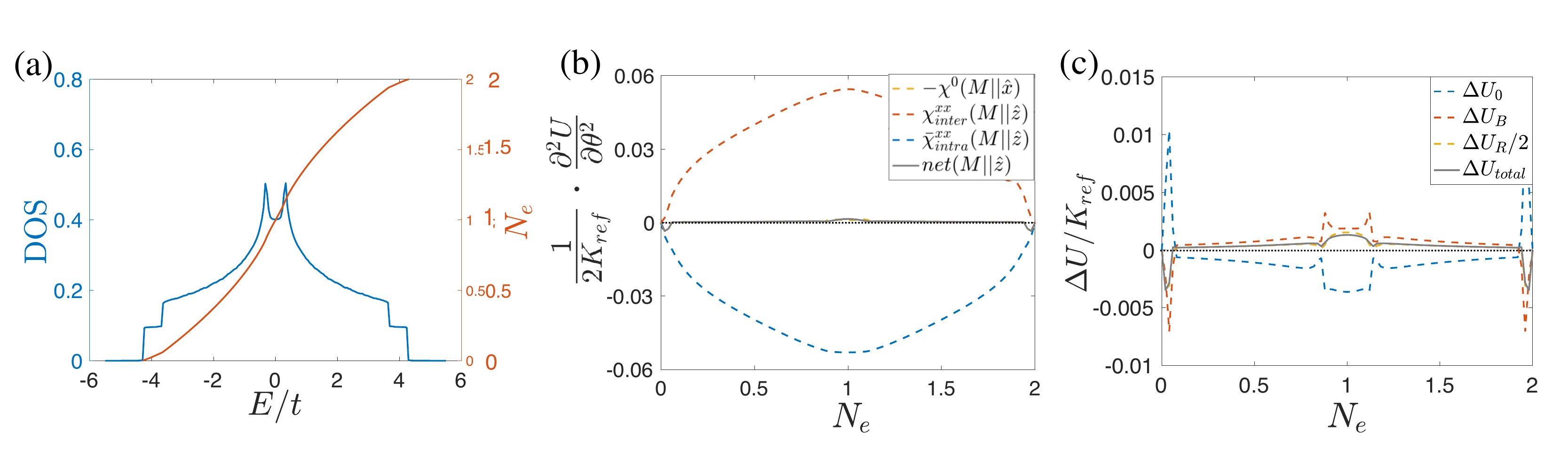}
  \caption{\label{fig:maecase3}
  MAE for the weak exchange case, $t' \gg J \sim t''$.
  (a) Total DOS and number of electrons as a function of energy, for $\VEC{M} \parallel \hat{\VEC{z}}$.
  (b) MAE from the second derivatives of the band energy, from the connection to its phenomenological form.
  For $\VEC{M} \parallel \hat{\VEC{x}}$ (Eq.~\ref{eq:d2udt2suscx}), only $\chi^0$ contributes (Eq.~\eqref{eq:suscvol}).
  For $\VEC{M} \parallel \hat{\VEC{z}}$ (Eq.~\ref{eq:d2udt2suscz}), $\chi^0$ does not contribute, and we plot the intraband (Eq.~\eqref{eq:chiintra}) and interband (Eq.~\eqref{eq:chiinter}) contributions from the uniform spin susceptibility, as well as the net result.
  (c) Internal energy differences $\Delta U = U(\nicefrac{\pi}{2},0) - U(0,0)$, decomposed using Eq.~\ref{eq:uintsplit}.
  The curve showing half of the difference in the SOC energy overlaps almost perfectly with the net internal energy differences, which in turn agrees very well with the results obtained from the susceptibility calculations, c.f.~panel (b).
  Parameters: $J = t''$ and $\PHIR=\pi/20$ ($t' = 2.0t$, $t'' = 0.3t$).
  }
\end{figure*}

Lastly we consider the case where the exchange energy is comparable to the SOC strength, by setting $J = t''$.
In this regime, the splitting between the two bands is small, as seen in Fig.~\ref{fig:maecase3}(a), as the bandwidth is mostly set by $t'$, and $t' \gg t'' \sim J$.
The intraband and interband contributions to the susceptibility are almost identical, Fig.~\ref{fig:maecase3}(b), leading to a small net value of the MAE.
Now we find PMA for almost all values of $N_{\text{e}}$, except at the band edges ($N_{\text{e}} \approx 0$ or 2) where IMA is recovered.
For these limiting values of the filling the band dispersions can be approximated by the free-electron Rashba model, for which IMA is the expected result~\cite{Barnes2014,Kim2016a}. Although it is not strictly applicable in this case, Eq.~\eqref{eq:k2approx} predicts that the range of $N_{\text{e}}$ around half-filling where PMA is found is expected to become wider as $J$ gets weaker, as observed in our data.
Fig.~\ref{fig:maecase3}(c) provides a better view of the behavior of the MAE, using the band energy difference between $\VEC{M} \parallel \hat{\VEC{x}}$ and $\VEC{M} \parallel \hat{\VEC{z}}$ and its decomposition.
As found for the previous case, when both bands are partially filled there is no direct correspondence between the contributions to the band energy difference and the contributions to the susceptibility.
Estimating the band energy difference by half of the anisotropy of the Rashba energy remains an excellent approximation, also in very good agreement with the results from the volume susceptibility.
\

%%%%%%%%%%%%%%%%%%%%%%%%%%%%%%%%%%%%%%%%%%%%%%%%%%%%%%%%%%%%%%%%%%%%%%%%%%%%%%%%%%%%%%%%%%%%%%%%%%%%
\section{Discussion and conclusions}\label{sec:conclusions}
%%%%%%%%%%%%%%%%%%%%%%%%%%%%%%%%%%%%%%%%%%%%%%%%%%%%%%%%%%%%%%%%%%%%%%%%%%%%%%%%%%%%%%%%%%%%%%%%%%%%
In this work, we explored a simple tight-binding model of spin-orbit-coupled electrons exchange-coupled to a background ferromagnetic order parameter, meant to abstract the essential electronic structure properties of the interface between a ferromagnetic layer and a heavy-metal layer.
The simplicity of the model made it attractive to consider different approaches to the calculation of the magnetic anisotropy energy: a global approach, based on band energy differences, and a local approach, based on the curvature of the energy for an equilibrium ferromagnetic orientation.
Besides reproducing the results of previous work~\cite{Barnes2014,Kim2016a}, by decomposing the spin susceptibility into intra and interband contributions and connecting them to the anisotropy of different energy terms in the Hamiltonian, we provide a detailed view on how the competition between in-plane and perpendicular magnetic 
anisotropies is settled.
Reassuringly, the global and local approaches to the magnetic anisotropy are found to be compatible, due to weak higher-order anisotropy contributions.
Perturbation theory was used to prove analytically that when the system is gapped, perpendicular magnetic anisotropy always ensues.

We found that the perpendicular magnetic anisotropy can be enhanced by tuning the splitting of the energy bands to the point where the gap between them is about to close (besides the obvious path of increasing the magnitude of the spin-orbit coupling).
This has the added advantage of increasing the range of filling values for which perpendicular magnetic anisotropy is present in the model.
The impact of tuning the effective splitting of the energy bands at the interface between a ferromagnet and a heavy-metal can be explored, both with density functional theory calculations and experimentally, by inserting dopants or a decoupling layer at the interface.
These studies would also uncover which are the generic features of the interface-driven magnetic anisotropy and which are the model-specific ones.

More importantly, we have shown that the magnetic anisotropy is usefully viewed 
as arising from a competition between Fermi surface and Fermi sea terms, 
with PMA arising when the former is overcome by the latter.
This should also hold for more complicated band structures whose details
 may also play an important role in determining how this competition is settled.
In Eq.~\eqref{eq:k2approx} we presented a simple approximate relation for the leading uniaxial anisotropy coefficient, $K_2 \approx K_{\text{ref}}\,\big(M - J \rho(\EF)\big)$,
The overall scale of the MAE is given by $K_{\text{ref}} = (t'')^2/4J$.
The interband term scales with the magnitude of the spin moment $M$, explaining why it is largest when the majority 
band is completely full and the minority band is completely empty.
The intraband term scales with the density of electronic states at the Fermi 
energy $\rho(\EF)$, and so is most important when the Fermi surface is large and bands are flat.
The competition between the two quantities is set by the magnitude of the exchange splitting $J$.
We speculate that these simple considerations should also extend to more complex multiband systems, either if the magnetic anisotropy is contributed mostly by a single pair of bands, or if the multiband spin susceptibility can be well-approximated by a sum of pairwise band contributions.
In this way, the electronic structure of the interface states can engineered in order to optimize PMA.

On the theoretical side, the calculation of the magnetic anisotropy energy from realistic band structures remains a challenging problem.
The magnetic force theorem has been employed to replace the total energy difference between two self-consistent calculations for orthogonal directions of the magnetization by the corresponding difference in band energies, requiring only one self-consistent calculation\cite{Daalderop1991}.
In a similar vein, the first derivative of the energy with respect to the orientation of the magnetization (the so-called magnetic torque) has also been effectively deployed\cite{Wang1996}.
Here we proposed to utilize the static uniform spin susceptibility to obtain the curvature of the energy for an equilibrium orientation of the magnetization, which requires a single self-consistent calculation.
We also validated the proposal of van der Laan~\cite{Laan1999} and Antropov~\cite{Antropov2014a} to consider the anisotropy of the spin-orbit coupling energy term in the Hamiltonian as an accurate approach to compute the magnetic anisotropy energy.
These two methods deserve further comparison within the context of realistic electronic structure calculations.
%%%%%%%%%%%%%%%%%%%%%%%%%%%%%%%%%%%%%%%%%%%%%%%%%%%%%%%%%%%%%%%%%%%%%%%%%%%%%%%%%%%%%%%%%%%%%%%%%%%%

%%%%%%%%%%%%%%%%%%%%%%%%%%%%%%%%%%%%%%%%%%%%%%%%%%%%%%%%%%%%%%%%%%%%%%%%%%%%%%%%%%%%%%%%%%%%%%%%%%%%
\acknowledgments
%%%%%%%%%%%%%%%%%%%%%%%%%%%%%%%%%%%%%%%%%%%%%%%%%%%%%%%%%%%%%%%%%%%%%%%%%%%%%%%%%%%%%%%%%%%%%%%%%%%%

G.C. acknowledges support from the Deutscher Akademischer Austauschdienst (DAAD) for a visit to
Forschungszentrum J\"ulich, where the initial part of the work was performed.  
M.d.S.D. and S.L. acknowledge funding by the European Research Council (ERC) under the European Union's Horizon 2020 research and innovation programme (ERC-Consolidator grant 681405 -- DYNASORE).
%%%%%%%%%%%%%%%%%%%%%%%%%%%%%%%%%%%%%%%%%%%%%%%%%%%%%%%%%%%%%%%%%%%%%%%%%%%%%%%%%%%%%%%%%%%%%%%%%%%%

%%%%%%%%%%%%%%%%%%%%%%%%%%%%%%%%%%%%%%%%%%%%%%%%%%%%%%%%%%%%%%%%%%%%%%%%%%%%%%%%%%%%%%%%%%%%%%%%%%%%
\appendix
\section{Internal energy vs. grand potential}\label{app:statphys}
%%%%%%%%%%%%%%%%%%%%%%%%%%%%%%%%%%%%%%%%%%%%%%%%%%%%%%%%%%%%%%%%%%%%%%%%%%%%%%%%%%%%%%%%%%%%%%%%%%%%
The properties of a system with a fixed number of electrons held at zero temperature can be derived from the internal energy, Eq.~\eqref{eq:intenergy}.
Suppose the hamiltonian depends on a set of parameters $\VEC{X}$, and we wish to find how the internal energy changes upon small changes in those parameters.
The first derivative is
\begin{align}
  \left.\frac{\partial U}{\partial X_i}\right|_{N_{\text{e}}} &= \left.\frac{\partial U}{\partial X_i}\right|_{\EF} + \frac{\partial U}{\partial \EF}\,\frac{\partial \EF}{\partial X_i}\notag\\
  &= \int^{\EF}_{-\infty}\hspace{-1em}\ud E\;\frac{\partial\rho(E,\VEC{X})}{\partial X_i}\,E
  + \rho(\EF,\VEC{X})\,\EF\,\frac{\partial\EF}{\partial X_i} \quad .
\end{align}
The vertical bars indicate which variables are kept fixed.
Using Eq.~\eqref{eq:elnumber} and the requirement of fixed number of electrons, its derivative must be zero,
\begin{align}\label{eq:fixednumber}
  0 &= \frac{\partial N_{\text{e}}}{\partial X_i} = \left.\frac{\partial N_{\text{e}}}{\partial X_i}\right|_{\EF} + \frac{\partial N_{\text{e}}}{\partial \EF}\,\frac{\partial \EF}{\partial X_i} \nonumber\\
  &= \int^{\EF}_{-\infty}\hspace{-1em}\ud E\;\frac{\partial\rho(E,\VEC{X})}{\partial X_i} + \rho(\EF,\VEC{X})\,\frac{\partial\EF}{\partial X_i} \quad,
\end{align}
so the first derivative of the internal energy can be rewritten as
\begin{equation}\label{eq:eintdiff1}
  \left.\frac{\partial U}{\partial X_i}\right|_{N_{\text{e}}} = \int^{\EF}_{-\infty}\hspace{-1em}\ud E\;\frac{\partial\rho(E,\VEC{X})}{\partial X_i}\,\big(E - \EF\big) \quad.
\end{equation}
This coincides with the first derivative of the grand potential,
\begin{equation}
  \Phi \underset{T=0}{=} U - \EF N_{\text{e}} \quad,\qquad \left.\frac{\partial \Phi}{\partial X_i}\right|_{\EF} = \left.\frac{\partial U}{\partial X_i}\right|_{N_{\text{e}}} \quad ,
\end{equation}
which is the expected thermodynamic result.
The grand canonical ensemble is often used instead of the canonical one, as calculations tend to be simpler.

Starting from Eq.~\eqref{eq:eintdiff1}, the second derivative of the internal energy is
\begin{align}
  \left.\frac{\partial^2 U}{\partial X_i\,\partial X_j}\right|_{N_{\text{e}}}  &= \int^{\EF}_{-\infty}\hspace{-1em}\ud E\;\frac{\partial^2\rho(E,\VEC{X})}{\partial X_i\,\partial X_j}\,\big(E - \EF\big)  \nonumber\\
  &\phantom{=} - \frac{\partial\EF}{\partial X_i}\int^{\EF}_{-\infty}\hspace{-1em}\ud E\;\frac{\partial\rho(E,\VEC{X})}{\partial X_j} \notag\\
  \label{eq:eintdiff2}
  &= \left.\frac{\partial^2\Phi}{\partial X_i\,\partial X_j}\right|_{\EF} + \rho(\EF,\VEC{X})\,\frac{\partial\EF}{\partial X_i}\,\frac{\partial\EF}{\partial X_j} \quad.
\end{align}
We see that the second derivatives are related by a factor which is related to how the number of electrons changes upon variation of the parameters in the hamiltonian.
This correction clearly vanishes for a gapped system ($\rho(\EF,\VEC{X}) = 0$) or when varying the parameters leaves the Fermi energy unchanged.

%%%%%%%%%%%%%%%%%%%%%%%%%%%%%%%%%%%%%%%%%%%%%%%%%%%%%%%%%%%%%%%%%%%%%%%%%%%%%%%%%%%%%%%%%%%%%%%%%%%%
\section{Green functions and the Hellmann-Feynman theorem}\label{app:gfhf}
%%%%%%%%%%%%%%%%%%%%%%%%%%%%%%%%%%%%%%%%%%%%%%%%%%%%%%%%%%%%%%%%%%%%%%%%%%%%%%%%%%%%%%%%%%%%%%%%%%%%
For our purposes, the Green function is the resolvent of the hamiltonian,
\begin{equation}
  \big(E - \MC{H}(\VEC{X})\big)\,G(E,\VEC{X}) = \mathcal{I} \quad,
\end{equation}
where the hamiltonian is assumed to depend on some parameters $\VEC{X}$, and $\mathcal{I}$ is the identity matrix for a chosen representation.
Taking the derivate with respect to the energy parameter we find
\begin{equation}
  \frac{\partial G(E,\VEC{X})}{\partial E} = -G(E,\VEC{X})\,G(E,\VEC{X}) \quad,
\end{equation}
and with respect to a hamiltonian parameter we get
\begin{equation}\label{eq:gfdiffparam}
  \frac{\partial G(E,\VEC{X})}{\partial X_i} = G(E,\VEC{X})\,\frac{\partial\MC{H}}{\partial X_i}\,G(E,\VEC{X}) \quad.
\end{equation}
Using the Dirac identity we obtain the spectral density matrix from the discontinuity of the Green function across the real energy axis,
\begin{align}\label{eq:imgf}
  \delta\big(E - \MC{H}(\VEC{X})\big) &= \lim_{\eta\rightarrow 0^+} \frac{G(E-\iu\eta,\VEC{X}) - G(E+\iu\eta,\VEC{X})}{2\pi\iu} \nonumber\\
  &\equiv -\frac{1}{\pi}\,\IM\,G(E,\VEC{X}) \quad .
\end{align}
The density of states of the system is then given by
\begin{equation}
  \rho(E,\VEC{X}) = -\frac{1}{\pi}\,\IM\,\Tr\,G(E,\VEC{X}) \quad ,
\end{equation}
and its derivative with respect to a hamiltonian parameter by
\begin{align}
  \frac{\partial\rho(E,\VEC{X})}{\partial X_i} &= -\frac{1}{\pi}\,\IM\,\Tr\,G(E,\VEC{X})\,\frac{\partial\MC{H}}{\partial X_i}\,G(E,\VEC{X}) \nonumber\\
   &= \frac{1}{\pi}\,\IM\,\Tr\,\frac{\partial G(E,\VEC{X})}{\partial E}\,\frac{\partial\MC{H}}{\partial X_i} \quad ,
\end{align}
using the cyclic property of the trace.

We can now replace these results in the first derivative of the internal energy, Eq.~\eqref{eq:eintdiff1},
\begin{align}
  \left.\frac{\partial U}{\partial X_i}\right|_{N_{\text{e}}} &= \frac{1}{\pi}\,\IM\,\Tr\int^{\EF}_{-\infty}\hspace{-1em}\ud E\;\frac{\partial G(E,\VEC{X})}{\partial E}\,\frac{\partial\MC{H}}{\partial X_i}\,\big(E - \EF\big) \notag\\
  \label{eq:hftheorem} 
  &= -\frac{1}{\pi}\,\IM\,\Tr\int^{\EF}_{-\infty}\hspace{-1em}\ud E\;G(E,\VEC{X})\,\frac{\partial\MC{H}}{\partial X_i} \nonumber\\
  &\equiv \left\langle\frac{\partial\MC{H}}{\partial X_i}\right\rangle \quad,
\end{align}
after integration by parts.
This is the Hellmann-Feynman theorem~\cite{Hellmann1937,Feynman1939}: the derivative of the energy with respect to a parameter is given by the ground state expectation value of the derivative of the hamiltonian with respect to the same parameter.

In Eq.~\eqref{eq:fixednumber} we find
\begin{align}
  \left.\frac{\partial N_{\text{e}}}{\partial X_i}\right|_{\EF} &= \int^{\EF}_{-\infty}\hspace{-1em}\ud E\;\frac{\partial\rho(E,\VEC{X})}{\partial X_i} \nonumber\\
  &= \frac{1}{\pi}\,\IM\,\Tr\,G(\EF,\VEC{X})\,\frac{\partial\MC{H}(\VEC{X})}{\partial X_i} \nonumber\\
  &\equiv -\left\langle\frac{\partial\MC{H}}{\partial X_i}\right\rangle_{\!\EF} \quad,
\end{align}
and using this and Eq.~\eqref{eq:gfdiffparam} we can express the second derivative of the internal energy, Eq.~\eqref{eq:eintdiff2}, as
\begin{align}\label{eq:eintdiff2gf}
  \left.\frac{\partial^2 U}{\partial X_i\,\partial X_j}\right|_{N_{\text{e}}} &= \left\langle\frac{\partial^2\MC{H}}{\partial X_i\,\partial X_j}\right\rangle \nonumber\\ 
  &- \frac{1}{\pi}\,\IM\,\Tr\!\int^{\EF}_{-\infty}\hspace{-1em}\ud E\;G(E)\,\frac{\partial\MC{H}}{\partial X_i}\,G(E)\,\frac{\partial\MC{H}}{\partial X_j} \nonumber\\
  &+ \frac{1}{\rho(\EF)}\;\left\langle\frac{\partial\MC{H}}{\partial X_i}\right\rangle_{\!\!\EF}\!\left\langle\frac{\partial\MC{H}}{\partial X_j}\right\rangle_{\!\!\EF} \quad.
\end{align}
The last term must be omitted for a gapped system (no Fermi surface).

%%%%%%%%%%%%%%%%%%%%%%%%%%%%%%%%%%%%%%%%%%%%%%%%%%%%%%%%%%%%%%%%%%%%%%%%%%%%%%%%%%%%%%%%%%%%%%%%%%%%
\section{Anatomy of the static uniform susceptibility}\label{app:anatomy}
%%%%%%%%%%%%%%%%%%%%%%%%%%%%%%%%%%%%%%%%%%%%%%%%%%%%%%%%%%%%%%%%%%%%%%%%%%%%%%%%%%%%%%%%%%%%%%%%%%%%
In this appendix the expression for the susceptibility using Green functions, Eq.~\eqref{eq:susc0} (see also Eq.~\eqref{eq:eintdiff2gf}), is recast in the more familiar form from perturbation theory.
We recall the spectral representation of the Green function, Eq.~\eqref{eq:gfspectral}:
\begin{align}
  G(\VEC{k},E) &= \sum_n \frac{P_n(\VEC{k})}{E - E_n(\VEC{k})} \quad,\nonumber\\
  P_n(\VEC{k}) &= \frac{1}{2}\left(\sigma^0 + n\,\hat{\VEC{b}}(\VEC{k})\cdot\VEC{\upsigma}\right) \quad,\qquad
  n = \pm \quad .
\end{align}
We only have to rewrite the term involving the product of Green functions,
%\begin{widetext}
\begin{align}
  %\left.\chi^{\alpha\beta}\right|_{\EF}\!
  &\phantom{=}\hspace{0.5em}\frac{1}{\pi}\,\IM\,\Tr\int^{\EF}_{-\infty}\hspace{-1em}\ud E\!\int\!\frac{\ud\VEC{k}}{(2\pi)^2}\;\sigma^\alpha\,G(\VEC{k},E)\,\sigma^\beta\,G(\VEC{k},E) \notag\\
  &= \sum_{n'n}\frac{1}{\pi}\,\IM\,\Tr\!\int^{\EF}_{-\infty}\!\hspace{-1em}\ud E\!\int\!\frac{\ud\VEC{k}}{(2\pi)^2}\;\frac{\sigma^\alpha P_{n'}(\VEC{k})}{E - E_{n'}(\VEC{k})}\,\frac{\sigma^\beta P_n(\VEC{k})}{E - E_n(\VEC{k})} \quad.
\end{align}
%\end{widetext}

To evaluate the energy integral we require the partial fraction decomposition of
\begin{align}
  &\phantom{=}\hspace{0.7em}\frac{1}{E - E_{n'}(\VEC{k})}\,\frac{1}{E - E_{n}(\VEC{k})} \nonumber\\
  &= \frac{1}{E_{n'}(\VEC{k}) - E_{n}(\VEC{k})}\left(\frac{1}{E - E_{n'}(\VEC{k})} - \frac{1}{E - E_{n}(\VEC{k})}\right) \quad,
\end{align}
which holds only if $n \neq n'$ (interband contribution), and contributes simple poles to the energy integral:
\begin{equation}
  -\frac{1}{\pi}\,\IM\,\Tr\int^{\EF}_{-\infty}\hspace{-1em}\ud E\;\frac{1}{E - E_{n}(\VEC{k})}
  \equiv f_n(\VEC{k}) \quad ,
\end{equation}
with $f_n(\VEC{k}) = \Theta\big(\EF - E_{n}(\VEC{k})\big)$ being the zero-temperature limit of the Fermi-Dirac distribution.
When $n' = n$ we have degeneracies (intraband term), which contribute a second-order pole and so have to be treated separately:
\begin{equation}
  -\frac{1}{\pi}\,\IM\,\Tr\int^{\EF}_{-\infty}\hspace{-1em}\ud E\;\frac{1}{\big(E - E_{n}(\VEC{k})\big)^2}
  = \frac{\partial f_n(\VEC{k})}{\partial E_{n}(\VEC{k})} \quad . %= -\delta\big(\EF - E_{n}(\VEC{k})\big)
\end{equation}
The matrix elements are given by
\begin{align}
  \MC{M}_{n'n}^{\alpha\beta}(\VEC{k}) &= \Tr\,\sigma^\alpha\,P_{n'}(\VEC{k})\,\sigma^\beta\,P_n(\VEC{k}) \notag\\
  \label{eq:melem}
  &= \frac{1 - n'n}{2}\,\delta_{\alpha\beta} + n'n\,\hat{b}_\alpha(\VEC{k})\,\hat{b}_\beta(\VEC{k}) \nonumber\\
  &- \iu\,\frac{n'-n}{2}\sum_\gamma \varepsilon_{\alpha\beta\gamma}\,\hat{b}_\gamma(\VEC{k}) \quad ,
\end{align}
with $\varepsilon_{\alpha\beta\gamma}$ the Levi-Civita symbol.
We can then write the susceptibility as $\chi^{\alpha\beta} = \chi^{\alpha\beta}_{\text{intra}} + \chi^{\alpha\beta}_{\text{inter}}$, with
\begin{align}\label{eq:suscintra}
  \chi^{\alpha\beta}_{\text{intra}}
  &= \!\int\!\frac{\ud\VEC{k}}{(2\pi)^2}\,\hat{b}_\alpha(\VEC{k})\,\hat{b}_\beta(\VEC{k}) \sum_{n=\pm} \delta\big(\EF - E_{n}(\VEC{k})\big) \nonumber\\
  &-\frac{m^\alpha(\EF)\,m^\beta(\EF)}{\rho(\EF)} \quad,
\end{align}
and
\begin{align}\label{eq:suscinter}
  \chi^{\alpha\beta}_{\text{inter}}
  = 2\!\int\!\frac{\ud\VEC{k}}{(2\pi)^2}\,\big(\hat{b}_\alpha(\VEC{k})\,\hat{b}_\beta(\VEC{k}) - \delta_{\alpha\beta}\big)\,
  \frac{f_{-}(\VEC{k}) - f_{+}(\VEC{k})}{E_{-}(\VEC{k})-E_{+}(\VEC{k})} \quad.
\end{align}
The intraband term collects the contributions from the Fermi energy, while the interband term collects those from the Fermi sea.
The contribution of the antisymmetric part of the matrix element to the interband term cancels out.

%%%%%%%%%%%%%%%%%%%%%%%%%%%%%%%%%%%%%%%%%%%%%%%%%%%%%%%%%%%%%%%%%%%%%%%%%%%%%%%%%%%%%%%%%%%%%%%%%%%%
\subsection{Ferromagnetic system without spin-orbit coupling}\label{app:anatomynosoc}
%%%%%%%%%%%%%%%%%%%%%%%%%%%%%%%%%%%%%%%%%%%%%%%%%%%%%%%%%%%%%%%%%%%%%%%%%%%%%%%%%%%%%%%%%%%%%%%%%%%%
For this example we can take $\hat{\VEC{b}}(\VEC{k}) = \VEC{S} = \hat{\VEC{z}}$ without loss of generality, as without SOC the system is invariant under spin rotations.
The energy dispersion of Eq.~\eqref{eq:bands} becomes
\begin{equation}
  E_n(\VEC{k}) = E_0(\VEC{k}) - n J \quad,
\end{equation}
and the matrix elements simplify to
\begin{align}
  \MC{M}_{n'n}^{\alpha\beta}(\VEC{k}) = \delta_{\alpha\beta}\,\frac{1 - n'n}{2} + n'n\,\delta_{\alpha z}\,\delta_{\beta z} - \iu\,\varepsilon_{\alpha\beta z}\,\frac{n'-n}{2} \quad.
\end{align}
The longitudinal susceptibility ($\alpha=\beta=z$) arises from the intraband contributions ($n' = n$), while the transverse susceptibility ($\alpha,\beta=x,y$) arises from the interband contributions ($n' \neq n$).
From Eq.~\eqref{eq:suscintra}, the longitudinal susceptibility is thus
\begin{align}
  \chi^{zz} &= \rho_+(\EF)+\rho_-(\EF) + \frac{\big(\rho_+(\EF)-\rho_-(\EF)\big)^2}{\rho(\EF)} \nonumber\\
  &= \frac{4\rho_+(\EF)\rho_-(\EF)}{\rho(\EF)} \quad,
\end{align}
with $\rho_n(\EF)$ the density of states at the Fermi energy of the $n$-band (check Eq.~\eqref{eq:dos}).
Here the correction term is crucial: if one band is partially occupied, $\rho_+(\EF) \neq 0$, and the other band is empty, $\rho_-(\EF) = 0$, then $\chi^{zz} = 0$, as the increase in the spin moment (the $\rho_+(\EF)$ contribution from the first term) is cancelled by the requirement of fixed number of electrons (enforced by the correction term).
From Eq.~\eqref{eq:suscinter}, the transverse susceptibility is (check Eq.~\eqref{eq:mspin})
\begin{align}
  \chi^{xx} = \chi^{yy}
  = -\!\int\!\frac{\ud\VEC{k}}{(2\pi)^2}\,\frac{f_{-}(\VEC{k}) - f_{+}(\VEC{k})}{J_{\text{sd}}}  = \frac{M}{J} \quad.
\end{align}
This cancels precisely the volume susceptibility, $\chi^0$, and makes the derivatives of the internal energy with respect to the angles defining the ferromagnetic direction vanish, Eqs.~\eqref{eq:d2udt2suscx}, \eqref{eq:d2udp2suscx} and \eqref{eq:d2udt2suscz}.
As discussed in Sec.~\ref{sec:mae}, this term is also present in the general case both in the transverse susceptibilities and in the volume susceptibility, and so those quantities are defined in the main text by analytically subtracting this term from both of them.

%%%%%%%%%%%%%%%%%%%%%%%%%%%%%%%%%%%%%%%%%%%%%%%%%%%%%%%%%%%%%%%%%%%%%%%%%%%%%%%%%%%%%%%%%%%%%%%%%%%%
\subsection{Gapped system at half-filling with $\VEC{S} = \hat{\VEC{z}}$}\label{app:anatomywithsoc}
%%%%%%%%%%%%%%%%%%%%%%%%%%%%%%%%%%%%%%%%%%%%%%%%%%%%%%%%%%%%%%%%%%%%%%%%%%%%%%%%%%%%%%%%%%%%%%%%%%%%
Now the Fermi energy lies in the gap, so one of the bands is fully occupied, $f_+(\VEC{k}) = 1$, and the other is empty, $f_-(\VEC{k}) = 0$.
Thus there are no intraband contributions to the susceptibility and the Fermi surface corrections vanish.
From Eq.~\eqref{eq:suscinter} and inserting the band dispersions of Eq.~\eqref{eq:bands}, the susceptibility is then
\begin{align}
  \chi^{\alpha\beta}
  = \!\int\!\frac{\ud\VEC{k}}{(2\pi)^2}\;\frac{\delta_{\alpha\beta} - \hat{b}_{\alpha}(\VEC{k})\,\hat{b}_{\beta}(\VEC{k})}{\NORM{\VEC{b}(\VEC{k})}} \quad.
\end{align}

The longitudinal susceptibility is
\begin{align}
  \chi^{zz}
  &= \!\int\!\frac{\ud\VEC{k}}{(2\pi)^2}\;\frac{1 - \big(\hat{b}_{z}(\VEC{k})\big)^2}{\NORM{\VEC{b}(\VEC{k})}} \nonumber\\
  &= \!\int\!\frac{\ud\VEC{k}}{(2\pi)^2}\;\frac{\big(\hat{b}_{x}(\VEC{k})\big)^2 + \big(\hat{b}_{y}(\VEC{k})\big)^2}{\NORM{\VEC{b}(\VEC{k})}} \neq 0 \quad,
\end{align}
which shows that the net spin moment is not saturated, due to SOC.
Using Eq.~\eqref{eq:spinaxis} we find ($a = \sin^2 k_x$ and $b = \sin^2 k_y$)
\begin{align}
  \chi^{zz}
  &= \!\int\!\frac{\ud\VEC{k}}{(2\pi)^2}\;\frac{(t'')^2 \left(a + b\right)}{\left((t'')^2 \left(a + b\right) + (J)^2\right)^{\frac{3}{2}}} 
  \notag\\
  %&= -t''\,\frac{\partial}{\partial t''}\!\int\!\frac{\ud\VEC{k}}{(2\pi)^2}\,\left((t'')^2 \left(a + b\right) + (J)^2\right)^{-\frac{1}{2}} \notag\\
  %&\approx \frac{t''}{J_{\text{sd}}}\,\frac{\partial}{\partial t''}\!\int\!\frac{\ud\VEC{k}}{(2\pi)^2}\,\left(\frac{(t'')^2}{2(J)^2} \left(a + b\right)
  %- \frac{3(t'')^4}{8(J)^4} \left(a + b\right)^2\right) \notag\\
  &\approx \frac{(t'')^2}{(J)^3} \!\int\!\frac{\ud\VEC{k}}{(2\pi)^2}\,\left(\left(a + b\right)
  - \frac{3(t'')^2}{2(J)^2} \left(a + b\right)^2\right) \notag\\
  &= \frac{(t'')^2}{(J)^3} \left(1 - \frac{15(t'')^2}{8(J)^2}\right) \quad.
\end{align}
The generating polynomial of Eq.~\eqref{eq:genpoly} was used to systematically evaluate the integrals.

The transverse susceptibility is
\begin{align}
  \chi^{xx} &= \!\int\!\frac{\ud\VEC{k}}{(2\pi)^2}\;\frac{1 - \big(\hat{b}_x(\VEC{k})\big)^2}{\NORM{\VEC{b}(\VEC{k})}} \nonumber\\
  &= \!\int\!\frac{\ud\VEC{k}}{(2\pi)^2}\;\frac{\big(\hat{b}_y(\VEC{k})\big)^2 + \big(\hat{b}_z(\VEC{k})\big)^2}{\NORM{\VEC{b}(\VEC{k})}} \notag\\
  &= \!\int\!\frac{\ud\VEC{k}}{(2\pi)^2}\;\frac{(t'')^2\,a + (J)^2}{\left((t'')^2\left(a + b\right) + (J)^2\right)^{\frac{3}{2}}} \notag\\
  &= \!\int\!\frac{\ud\VEC{k}}{(2\pi)^2}\;\frac{\frac{(t'')^2}{2}\left(a + b\right) + (J)^2}{\left((t'')^2\left(a + b\right) + (J)^2\right)^{\frac{3}{2}}}  \quad. %\notag\\
  %&= -\left(\frac{t''}{2}\,\frac{\partial}{\partial t''} + J\,\frac{\partial}{\partial J}\right)\!\int\!\frac{\ud\VEC{k}}{(2\pi)^2}\,\left((t'')^2\left(a + b\right) + (J)^2\right)^{-\frac{1}{2}} \quad.
\end{align}
The symmetry of the integrand allows the replacement shown on the third line, and in turn shows that $\chi^{yy} = \chi^{xx}$.
Substituting the results from the longitudinal susceptibility,
\begin{align}
  \chi^{xx} &= \frac{1}{2}\,\chi^{zz} - J\,\frac{\partial\chi^0}{\partial J} \nonumber\\
  &= \frac{1}{2}\,\chi^{zz} - \frac{\partial M}{\partial J} + \chi^0 \nonumber\\
  &= -\frac{1}{2}\,\chi^{zz} + \chi^0 \quad.
\end{align}
These identifications follow from the expression for the longitudinal uniform susceptibility and from the definition of the spin moment, Eq.~\eqref{eq:mspin}, which under the present assumptions leads to the volume susceptibility
\begin{equation}
  \chi^0 \equiv \frac{M}{J} = \!\int\!\frac{\ud\VEC{k}}{(2\pi)^2}\;\frac{\hat{b}_z(\VEC{k})}{J}
  = \!\int\!\frac{\ud\VEC{k}}{(2\pi)^2}\;\frac{1}{\NORM{\VEC{b}(\VEC{k})}} \quad .
\end{equation}
Using the expansion found for the longitudinal susceptibility,
\begin{align}
  \chi^0 &\approx \!\int\!\frac{\ud\VEC{k}}{(2\pi)^2}\,\left(\frac{(t'')^2}{2(J)^2}\left(a + b\right)
   - \frac{3(t'')^4}{8(J)^4}\left(a + b\right)^2\right) \notag\\
  &= \frac{(t'')^2}{2(J)^2}\left(1 - \frac{15(t'')^2}{16(J)^2}\right) \quad,
\end{align}
and from Eq.~\eqref{eq:d2udt2suscz} we obtain the uniaxial magnetic anisotropy coefficient, 
\begin{align}
  \left.\frac{\partial^2 U}{\partial\theta^2}\right|_{\VEC{M} \parallel \hat{\VEC{z}}} &= \frac{1}{2}\,(J)^2\,\chi^{zz} \nonumber\\
  &= \frac{1}{2}\,\frac{(t'')^2}{J} - \frac{15}{16}\,\frac{(t'')^4}{(J)^3}
  = 2 K_2 \quad,
\end{align}
in perfect agreement with the direct calculation of Sec.~\ref{sec:halffilling}.
%%%%%%%%%%%%%%%%%%%%%%%%%%%%%%%%%%%%%%%%%%%%%%%%%%%%%%%%%%%%%%%%%%%%%%%%%%%%%%%%%%%%%%%%%%%%%%%%%%%%

%%%%%%%%%%%%%%%%%%%%%%%%%%%%%%%%%%%%%%%%%%%%%%%%%%%%%%%%%%%%%%%%%%%%%%%%%%%%%%%%%%%%%%%%%%%%%%%%%%%%
\bibliography{bibliography}
%%%%%%%%%%%%%%%%%%%%%%%%%%%%%%%%%%%%%%%%%%%%%%%%%%%%%%%%%%%%%%%%%%%%%%%%%%%%%%%%%%%%%%%%%%%%%%%%%%%%

\end{document}